\documentclass[aps,prb,twocolumn,superscriptaddress]{revtex4-2}
\usepackage{amsmath, amssymb,  graphicx,bbm}

\usepackage[colorlinks=true ,urlcolor=blue,urlbordercolor={0 1 1}]{hyperref}

\usepackage{color}
\usepackage{xcolor}
\usepackage{braket}
\usepackage{natbib}
\usepackage[utf8]{inputenc}
\usepackage{bm}
\usepackage{verbatim}
\usepackage{graphicx}
\usepackage{amsmath}
\usepackage{color}
\usepackage{float}
\usepackage{bbm}
\usepackage{amssymb}
\usepackage{slashed}
\usepackage{wasysym,bm,bbm,dsfont,braket}

\begin{document}
\title{Topological incommensurate Fulde–Ferrell–Larkin–Ovchinnikov  superconductor and Bogoliubov Fermi surface in rhombohedral tetra-layer graphene}
\author{Hui Yang}
\author{Ya-Hui Zhang}
\affiliation{William H. Miller III Department of Physics and Astronomy, Johns Hopkins University, Baltimore, Maryland, 21218, USA}
\begin{abstract}
    We performed a random phase approximation (RPA) calculation for a spin-valley polarized model of the rhombohedral tetra-layer graphene to study the possibility of chiral superconductor from the Kohn-Luttinger mechanism. We included the realistic band structure and form factor in our calculation and solved the self-consistent equation numerically by sampling 20,000 points in the momentum space at a given temperature. Around the Van-Hove singularity (VHS), we find p-ip pairing with Chern number switching from $C=-1$ to $C=0$ through a gap closing at $\mathbf k=(0,0)$ (defined relative to $\mathbf K$). Although the superconductor is generically fully gapped at low temperature, we find Bogoliubov Fermi surface at temperature just below mean field $T_c$. Besides, through calculation of the free energy, we conclude that the optimal Cooper pair momentum $\mathbf Q$ is generically finite and can be as large as $0.1 k_F$.  We dub the $\mathbf Q\neq 0$ phase as an incommensurate Fulde-Ferrell-Larkin-Ovchinnikov(FFLO) superconductor to distinguish it from the $\mathbf Q=0$ phase.  Compared to the $\mathbf Q=0$ phase, our incommensurate $\mathbf Q$ phase is a nematic superconductor if it is in the Fulde-Ferrell(FF) phase or exhibts charge density wave (CDW) if it is in the Larkin-Ovchinnikov (LO) phase. Our work demonstrates the rhombohedral tetra-layer graphene as a wonderful platform to explore Majorana zero-mode, FFLO physics and Bogoliubov fermi surface within one single platform.
\end{abstract}
\maketitle
\section{Introduction}
There have been lots of efforts on searching for a topological superconductor from higher angular momentum pairing\cite{RevModPhys.63.239}.  Especially  it is desirable to realize a spinless p+ip superconductor which hosts Majorana zero mode and may be useful for the fault-tolerant quantum computation\cite{Alicea_2012,Nayak2008Non,ZhangRMP,RevModPhys.82.3045,Read2000paired,Schnyder2008Classification, Nayak2008Non,Kitaev2009Periodic,Sato2017topological}.  Despite many attempts in the past decades, well-established evidence of a topological and chiral superconductor remains elusive. Signatures of chiral superconducting phase in Sr$_2$RuO$_4$\cite{Maeno1994}and UTe$_2$\cite{doi:10.7566/JPSJNC.16.08,Jiao2020,Aoki_2022} were reported, but the pairing symmetry is still under debate\cite{Theuss2024,doi:10.1073/pnas.2020492118,Pustogow2019,PhysRevB.104.024511}.

The recent progress in the two dimensional materials provides wonderful platforms for various exotic topological phases\cite{Andrei2021,Nuckolls2024,Balents2020,Cao2018Unconventional,Yankowitz2019Tuning,Lu2019Superconductors,Han2024signatures,choi2024electric,Xia2024,guo2024superconductivity,Jindal2023,jia2024anomalous}. Especially superconductivity has been reported in several systems now. The list includes the magic-angle twisted bilayer graphene\cite{Cao2018Unconventional,doi:10.1126/science.aav1910,Arora2020,Cao2021,Oh2021} (TBG) and twisted trilayer graphene (TTG)\cite{Park2021,doi:10.1126/science.abg0399}, moir\'eless bilayer\cite{doi:10.1126/science.abm8386,Zhang2023,Li2024,holleis2023ising,zhang2024twist}, trilayer\cite{Zhou2021,patterson2024superconductivity,yang2024diverse} and tetralayer graphene\cite{choi2024electric,Han2024signatures}, and twisted WSe$_2$\cite{Xia2024,guo2024superconductivity}. These experimental observations have also led to many theoretical proposals for the possible mechanism of superconductivity, including phonon-driven pairing\cite{PhysRevLett.121.257001,PhysRevB.99.165112,PhysRevLett.122.257002,PhysRevLett.127.187001,tuo2024theory,PhysRevB.106.024507,PhysRevB.105.L100503,PhysRevB.110.L041108}, weak coupling theory based on van-Hove singularity\cite{PhysRevX.8.041041,Yuan2019,PhysRevB.106.155115},  Kohn-Luttinger mechanism\cite{PhysRevLett.127.247001,PhysRevB.105.075432,PhysRevB.108.045404,PhysRevB.107.L161106,guerci2024topological,qin2024kohn}, pairing mediated by isospin fluctuations\cite{PhysRevLett.121.087001,You2019,Lee2019,Chatterjee2022,PhysRevB.105.L081407,PhysRevX.12.021018,PhysRevLett.127.217001,dong2023superconductivity,dong2024superconductivity}, slave boson theory\cite{PhysRevB.108.155111,kim2024theory} and more exotic scenarios based on topological defects\cite{doi:10.1126/sciadv.abf5299} and even anyons\cite{shi2024doping,Kim2024Topological}.

A recent experiment reported a possible chiral superconductor with $T_c\approx 300mK$ in tetra-layer rhombohedral graphene\cite{Han2024signatures}. The superconductor seems to emerge within a spin-valley polarized quarter metal phase. If true, a natural candidate is a topological p-wave superconductor.  Motivated by the experiment, we ask which pairing mechanism is possible in a purely spin-valley polarized model of the tetra-layer graphene. Now the options are quite limited. Isospin fluctuation scenario can be ruled out because we can always work in the maximally polarized sector where there is no fluctuation of isospin. Meanwhile it is not clear how phonon-driven pairing works for a spinless odd angular momentum channel. At least in the experiment there is no proximate fractional phase, so exotic anyon superconductivity does not seem to be likely for the realistic system. In the end we conclude that the most likely mechanism is the Kohn-Luttinger  superconductor from purely Coulomb interaction.

We performed the standard random phase approximation (RPA) calculation to incorporate the Kohn-Luttinger mechanism. The RPA theory has already been discussed by two previous works on the same system\cite{Chou2024intravalley,Geier2024Chiral}.   However, there are some simplifications in the previous calculations. For example, one key question is whether pairing is still possible when there is no inversion symmetry to guarantee the nesting condition $\epsilon(\mathbf k)=\epsilon(-\mathbf k)$. To answer this question, we need to include the trigonal warping term which is absent in Ref.~\cite{Geier2024Chiral}.  In this work we provide a full RPA calculation including realistic band structure and form factors in the interaction. Because there is no nesting condition now, the optimal total momentum $\mathbf Q$ of the Cooper pair is not necessarily zero (zero momentum is defined as the K corner of  the graphene Brillouin zone). Therefore we also target ansatz with non-zero $\mathbf Q$ and select the one with the lowest free energy. 

We provide a phase diagram in terms of the density n and the displacement field D for a single-flavor model in the valley K. We find $p-ip$ pairing with mean field $T_c \sim 2$ K\footnote{The real Berezinskii-Kosterlitz-Thouless (BKT) critical temperature is different from mean field $T_c$ and is decided by phase stiffness.} around the Van-Hove singularity (VHS). Because the time-reversal symmetry is already broken at the starting point, $p+ip$ pairing is not degenerate and is at higher energy. Contrary to the previous works, we find that the optimal $\mathbf Q$ is generically not zero and can reach $10\%$ of $k_F$. We note that even the $\mathbf Q=0$ phase can be viewed as a Fulde-Ferrell-Larkin-Ovchinnikov\cite{PhysRev.135.A550,Larkin:1964wok,RevModPhys.76.263} (FFLO) superconductor as we define the momentum relative to the $K$ corner of the graphene Brillouin zone. However, there is no breaking of the $C_3$ symmetry and there is no charge density wave (CDW) order. This is very different from the familiar FFLO with incommensurate momentum. In contrast, our $\mathbf Q\neq 0$ phase is an incommensurate FFLO superconductor and needs to break either the $C_3$ symmetry or exhibits a CDW order depending on whether it is in the FF or LO phase, in agreement with the usual FFLO phase.  For simplicity we restrict to the FF ansatz.  We find the spectrum of the Bogoliubov quasi particles does not change much for ansatz with $\mathbf Q=0$ and $\mathbf Q\neq 0$. At low temperature, in a large parameter region we have a fully gapped spectrum despite the lack of the nesting condition.  The Chern number switches from $C=-1$ to $C=0$ from increasing the displacement field $D$. Therefore we confirm the existence of a topological incommensurate FFLO superconductor hosting single Majorana zero mode in its vortex. At higher temperature, the Bogoliubov quasi particle gap can close and Bogoliubov Fermi surfaces emerge. In summary, our work demonstrates that the tetra-layer graphene system is a wonderful platform to study not only  fully gapped $p$-wave chiral superconductor, but also incommensurate FFLO physics and Bogoliubov Fermi surfaces. The later two aspects, to our best knowledge, have not been pointed out in previous theories.

\section{The Model}
\begin{figure}[htbp]
    \centering
\includegraphics[width=0.5\textwidth]{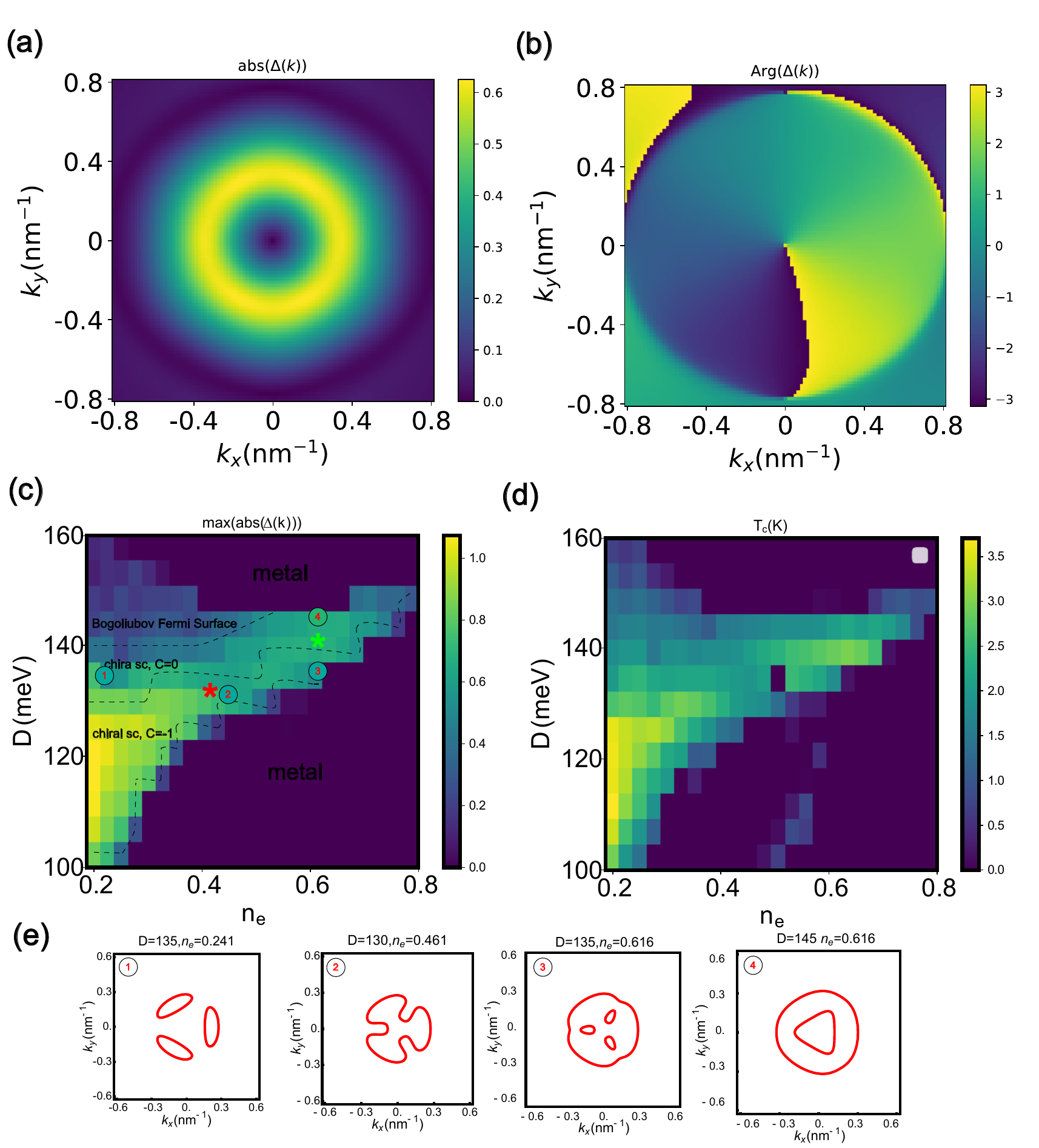}
    \caption{(a), (b) the amplitude and the phase of the order parameter $\Delta({\bf k})$ at ${\bf Q}=0$  at $(D,n_e)=(135,0.51)$, $\epsilon=6$, and $T=0$. (c) phase diagram of tetra-layer graphene at ${\bf Q}=(0,0)$ for $\epsilon=6$ and $d=20nm$ at $T=0$. We find three different superconducting phases at $T=0$: a topological superconductor with $C=-1$, a trivial chiral superconductor phase with $C=0$, and a gapless superconductor with Bogoliubov Fermi surfaces. (d) the critical temperature $T_c$ at $\epsilon=6$. (e) Illustration of the normal state Fermi surface correspond to parameters labeled as 1,2,3,4 in (c). }
    \label{fig:phase_diagram}
\end{figure}

We assume full spin-valley polarization. Focusing on one spin species of valley $K$, the  rhombohedral tetra-layer graphene is described by the following Hamiltonian\cite{min2008electronic}: 
\begin{align}
    H=H_0+H_{int},
    \label{eq:model}
\end{align}
where $H_0=\sum_{k}\Psi_{k}^\dag h_0(k)\Psi_k$ is the free part, $\Psi_k^\dag$ is the 8-dimensional operator defined in the layer and sublattice space and $h_0(k)$ is defined as
\begin{align}
\begin{pmatrix}
    \delta-\frac{D}{2}&v_0\Pi^\dagger&v_4\Pi^\dagger&v_3\Pi&0&\frac{\gamma_2}{2}&0&0\\
    v_0\Pi&-\frac{D}{2}&\gamma_1&v_4\Pi^\dagger&0&0&0&0\\
    v_4\Pi&\gamma_1&-\frac{D}{6}&v_0\Pi^\dagger&v_4\Pi^\dagger&v_3\Pi&0&\frac{\gamma_2}{2}\\
    v_3\Pi^\dagger&v_4\Pi&v_0\Pi&-\frac{D}{6}&\gamma_1&v_4\Pi^\dagger&0&0\\
    0&0&v_4\Pi&\gamma_1&\frac{D}{6}&v_0\Pi^\dagger&v_4\Pi^\dagger&v_3\Pi\\
    \frac{\gamma_2}{2}&0&v_3\Pi^\dagger&v_4\Pi&v_0\Pi&\frac{D}{6}&\gamma_1&v_4\Pi^\dagger\\
    0&0&0&0&v_4\Pi&\gamma_1&\frac{D}{2}&v_0\Pi^\dagger\\
    0&0&\frac{\gamma_2}{2}&0&v_3\Pi^\dagger&v_4\Pi&v_0\Pi&\delta+\frac{D}{2}
\end{pmatrix},
\label{eq:free_model}
\end{align}
where $\Pi=\tau k_x+ik_y$ ($\tau=\pm1$ corresponds to valley $K$ and $K^\prime$, respectively). $\mathbf k$ is defined relative to $\mathbf K$, the corner of the graphene Brillouin zone.  $v_j=\frac{\sqrt{3}}{2}\gamma_j a_0$, $\gamma_j$ is the velocity and $a_0=0.246nm$ is the lattice constant of graphene.  $\gamma_0=3100$meV, $\gamma_1=380$meV, $\gamma_2=-15$ meV, $\gamma_3=-290$meV, $\gamma_4=-141$meV, $\delta=-10.5$meV\cite{Zhou2021,Chou2024intravalley,PhysRevB.107.104502}. $D$ is the potential difference from a displacement field. As the displacement field $D$ and the electron doping are changed, the Fermi surface shape is changed with the emergence of van Hove singularities, as shown in Fig.~\ref{fig:phase_diagram}(e). 

The interaction is $H_{int}=\frac{1}{2A}\sum_{q}V_{\bf q}:\rho_{\bf q}\rho_{-{\bf q}}:$, where $A$ is the area of the sample, $V_{\bf q}=\frac{e^2}{2\epsilon\epsilon_0}\frac{\tanh{qd}}{q}$ is the Coulomb interaction screened by the distance between the gate and the sample. We fix $d=20$ nm in this work. We project the interaction into the first conduction band by substituting the density operator $\rho_{\bf q}$ with $\tilde{\rho}_{\bf q}=\sum_{\bf k}\Lambda_{{\bf k},{\bf q}}c^\dagger_{\bf k} c_{{\bf k}+{\bf q}}$, where $\Lambda_{{\bf k},{\bf q}}=\langle u_{\bf k}|u_{{\bf k}+{\bf q}}\rangle$ is the form factor, $c_{\bf k}^\dagger$ is the creation operator with momentum ${\bf k}$ in the first conduction band and $u_{\bf k}$ is the corresponding Bloch wave function.

\section{Random phase approximation and Kohn-Luttinger mechanism of superconductivity}

We perform a random phase approximation (RPA) to screen the interaction by the particle-hole excitations.  The RPA interaction at zero-frequency is given by
\begin{align}
    V^{RPA}_{\bf q}=\frac{V_{\bf q}}{1+V_{\bf q}\Pi_{\bf q}},
\end{align}
where $\Pi_q=\frac{1}{A}\sum_{\bf k}|\Lambda_{{\bf k},{\bf q}}|^2\frac{n_F(\epsilon_{{\bf k}+{\bf q}})-n_F(\epsilon_{\bf k})}{\epsilon_{\bf k}-\epsilon_{{\bf k}+{\bf q}}}$ is the polarization function, and $\epsilon_{\bf k}$ is the free dispersion of the first conduction band. $n_F(\epsilon)$ is the fermi distribution function. The renormalized interaction term $H_{r;int}$ becomes,
\begin{align}
    H_{r;int}=\frac{1}{2A}\sum_{{\bf k}_1,{\bf k}_2,{\bf q}}V^{RPA}_{{\bf q}}\Lambda_{{\bf k}_1,{\bf q}}\Lambda_{{\bf k}_2,-{\bf q}} c_{{\bf k}_1}^\dagger c_{{\bf k}_1+{\bf q}}c_{{\bf k}_2}^\dagger c_{{\bf k}_2-{\bf q}}.
    \label{eq:self_consistent}
\end{align}
We consider two electrons on the Fermi surface and calculate the interaction in the channel with angular momentum $l$, $V_l=\int d\theta V^{RPA}_{{\bf k}_1-{\bf k}_2}e^{i l\theta}$, where $\theta$ is the angle between ${\bf k}_1$ and ${\bf k}_2$  We find that $V_l$ is negative for $l=\pm1$  in a wide range of parameters, implying a pairing instability in the $p$-wave channel. To capture the superconducting order parameter, we perform a self-consistent calculation with the RPA interaction.

We introduce the superconducting order parameter
\begin{align}
    \Delta_{\bf Q}({\bf k}_1)=\frac{1}{A}\sum_{{\bf k}_2}\tilde{V}({\bf k}_2-{\bf k}_1)\langle c_{-{\bf k}_2+\frac{\bf Q}{2}}c_{{\bf k}_2+\frac{\bf Q}{2}}\rangle,
\end{align}
with $\tilde{V}({\bf k}_2-{\bf k}_1)=V({\bf k}_2-{\bf k}_1)\Lambda_{{\bf k}_1+\frac{\bf Q}{2},{\bf k}_2-{\bf k}_1}\Lambda_{-{\bf k}_1+\frac{\bf Q}{2},{\bf k}_1-{\bf k}_2}$. Here we also include the possibility of finite-momentum pairing with momentum ${\bf Q}$. The Bogoliubov-de-Gennes (BdG) Hamiltonian  reads
\begin{align}
    H_{BdG}=\sum_{{\bf k}\in BZ^\prime}
        \Psi_{\bf Q}^\dagger({\bf k})
    \begin{pmatrix}
        \epsilon_{k+\frac{Q}{2}}-\mu&\Delta_{\bf Q}({\bf k})\\ \Delta_{\bf Q}^*({\bf k})&-(\epsilon_{-k+\frac{Q}{2}}-\mu)
    \end{pmatrix}
    \Psi_{\bf Q}({\bf k})
\end{align}
where $\Psi_{\bf Q}({\bf k})=(c_{{\bf k}+\frac{{\bf Q}}{2}},c^\dagger_{-{\bf k}+\frac{{\bf Q}}{2}})^{T}$ is the Numbu spinor and ${\bf k}\in BZ^\prime$ means we sum over half the Brillouin zone with $k_x>0$. $\mu$ is the chemical potential introduced to fix the particle number.  From the BdG Hamiltonian, we can get the expectation value
\begin{align}
    \langle c_{-{\bf k}+\frac{\bf Q}{2}}c_{{\bf k}+\frac{\bf Q}{2}}\rangle=\frac{\Delta_{\bf k}({\bf k})}{2\eta_{k}}(\frac{1}{e^{\beta(e_0({\bf k})+\eta_{\bf k})}+1}-\frac{1}{e^{\beta(e_0({\bf k})-\eta_{\bf k})}+1}),
    \label{ansatz}
\end{align}
where $\beta=1/T$ is the inverse of temperature $T$, $e_0({\bf k})=\frac{1}{2}(\epsilon_{{\bf k}+\frac{\bf Q}{2}}-\epsilon_{-{\bf k}+\frac{\bf Q}{2}})$, $\eta({\bf k})=\sqrt{e_1^2({\bf k})+|\Delta({\bf k})|^2}$, with $e_1({\bf k})=\frac{1}{2}(\epsilon_{{\bf k}+\frac{\bf Q}{2}}+\epsilon_{-{\bf k}+\frac{\bf Q}{2}}-2\mu)$. In our RPA calculation, we use the meshgrid with $1000\times 1000$ momentum  points in the range $ka_0\in[-0.2,0.2]$ in each direction. We solve the self-consistent equation  by sampling around $20,000$ points in the momentum space with a dense meshgrid near the Fermi surface. We perform iterations  from a random ansatz $\Delta_{\bf Q}({\bf k})$ (not restricted to $p+ip$) at a fixed temperature. Because the Fermi energy is small compared to the interaction, we notice that $\Delta(\mathbf k)$ is not just confined around the Fermi surface, so the usual linearization procedure just along the Fermi surface may not be sufficient.

\section{phase diagram  with $\mathbf Q=(0,0)$ and topological phase transition}

We  first restrict to $\mathbf Q=(0,0)$ for simplicity. Later we will show that generically a small non-zero $\mathbf Q$ is favored, but the property is quite close to $\mathbf Q=0$. Thus calculation with $\mathbf Q=0$ is useful to quickly obtain a phase diagram. The amplitude and phase of the order parameter $\Delta({\bf k})$ at $(D,n_e)=(135,0.51)$ are shown in Fig.~\ref{fig:phase_diagram}(a) and (b). The phase of order parameter winds by $-2\pi$, indicating a $p-ip$ pairing.  Note that in the current model without time reversal symmetry, $p+ip$ is not degenerate with $p-ip$.  Moreover, we find the order parameter is non-zero in a wide range of momentum. We can get strong pairing along a stripe in the parameter space of electron density $n$ and displacement field $D$, as shown in Fig.~\ref{fig:phase_diagram}(c). Along the stripe we also confirm that $T_c$ can reach as large as $3$K.

At $T=0$, we confirm that most of the region with pairing has fully gapped quasi-particle spectrum. Close to the superconductor-metal transition there is a region with Bogoliubov Fermi surfaces because $\Delta(k)$ is not large enough to fully gap out the Fermi surface without the nesting condition $\epsilon(\mathbf k)=\epsilon(-\mathbf k)$ (see Fig.~\ref{fig:phase_diagram}(e) for Fermi surfaces at several representative points.)

For the fully gapped superconductor, we calculate the Chern number by integrating the Berry curvature in the whole Brillouin zone of the graphene, with contributions from both the valley $K$ and the valley $K'$.  For valley $K'$, we assume that there is no pairing and thus the Berry curvature is purely from the Bloch wave function. For the valley $K$, we have contributions from both the Bloch wave function and the $p-ip$ pairing.  In the end, the Chern number is 
\begin{align}
C=\frac{1}{2\pi}\int d^2{\bf k}B_+(k_x,k_y)+B^0_-(k_x,k_y),
\end{align}
where $B_+(k_x,k_y)$ is the Berry curvature in the valley K with $p-ip$ pairing, $B_-^0(k_x,k_y)$ is the bare Berry curvature from the Bloch wave function of the other valley $K^\prime$. We use the formula  $B^0_-({\bf k})=-B_+^0(-{\bf k})$ to obtain $B^0_-(\mathbf k)$.  We find that both $C=-1$ and $C=0$ are possible and there is a topological transition tuned by $D$.

\begin{figure}[htbp]
    \centering
\includegraphics[width=0.5\textwidth]{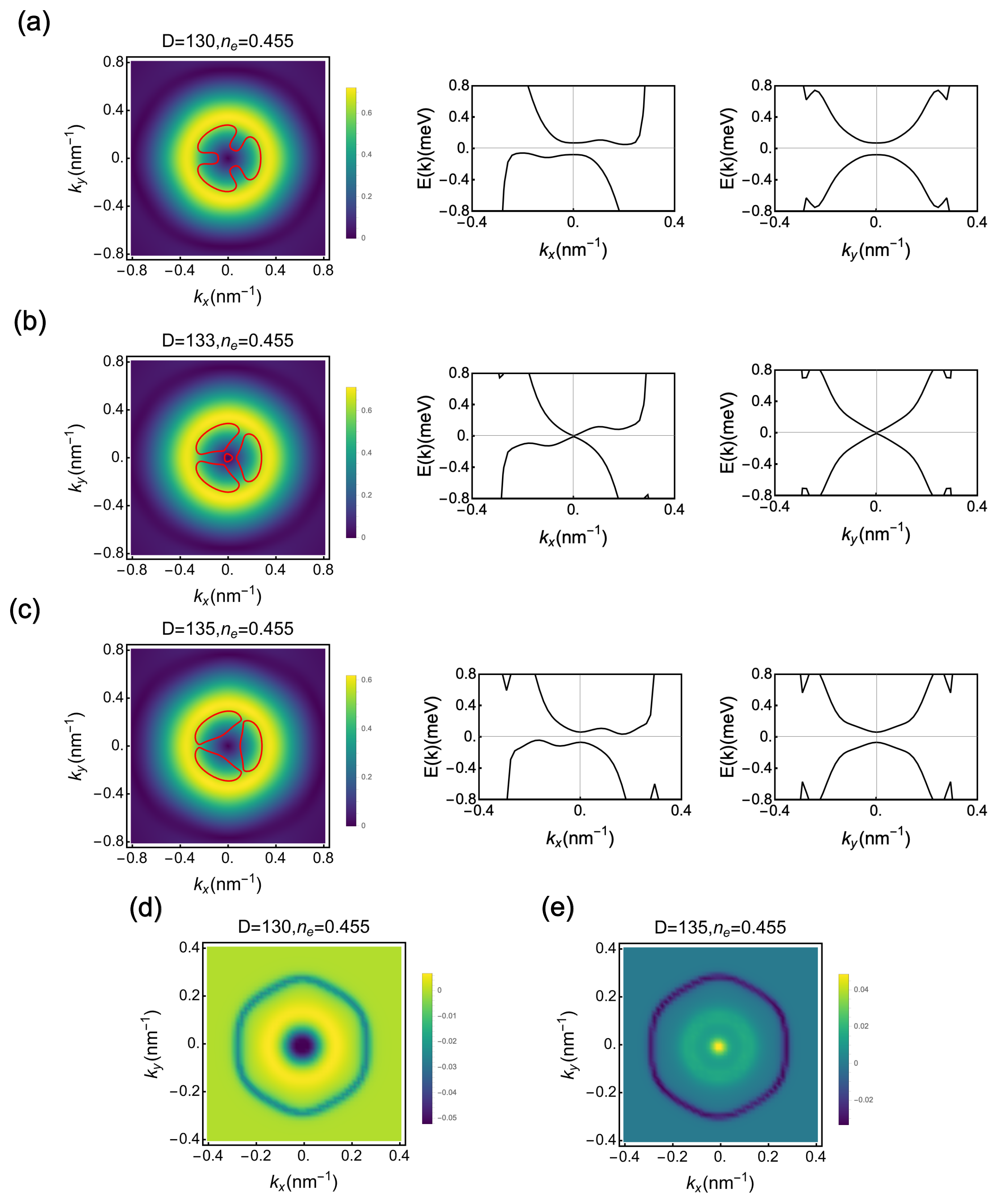}
    \caption{Topological phase transition for fixed electron density at $\epsilon=6$ and $T=0$. (a) topological superconductor with $C=-1$, (b) critical point with gap closing, (c) trivial gapped superconductor with $C=0$. The red lines correspond to the Fermi surfaces in normal state. (d), (e) correspond to the Berry curvature distributions of the topological superconductivity and trivial superconductivity, respectively. One can see that the topological transition is associated with a van-Hove singularity of the normal state. Within the superconductor phase, there is a gap closing at $\mathbf k=(0,0)$.}
    \label{fig:phase_transition}
\end{figure}

In Fig.~\ref{fig:phase_transition} we provide details on the topological phase transition tuned by D at a fixed density. The transition happens at the Lifshitz transition where the shape of Fermi surface changes. At the transition point, there is one single gapless Dirac cone at $\mathbf k=(0,0)$. The two sides correspond to different sign of the mass of this Dirac cone, resulting different sign of Berry curvature at $\mathbf k=(0,0)$ (see Fig.~\ref{fig:phase_transition}(d)(e)).

\begin{figure}[htbp]
    \centering
\includegraphics[width=0.5\textwidth]{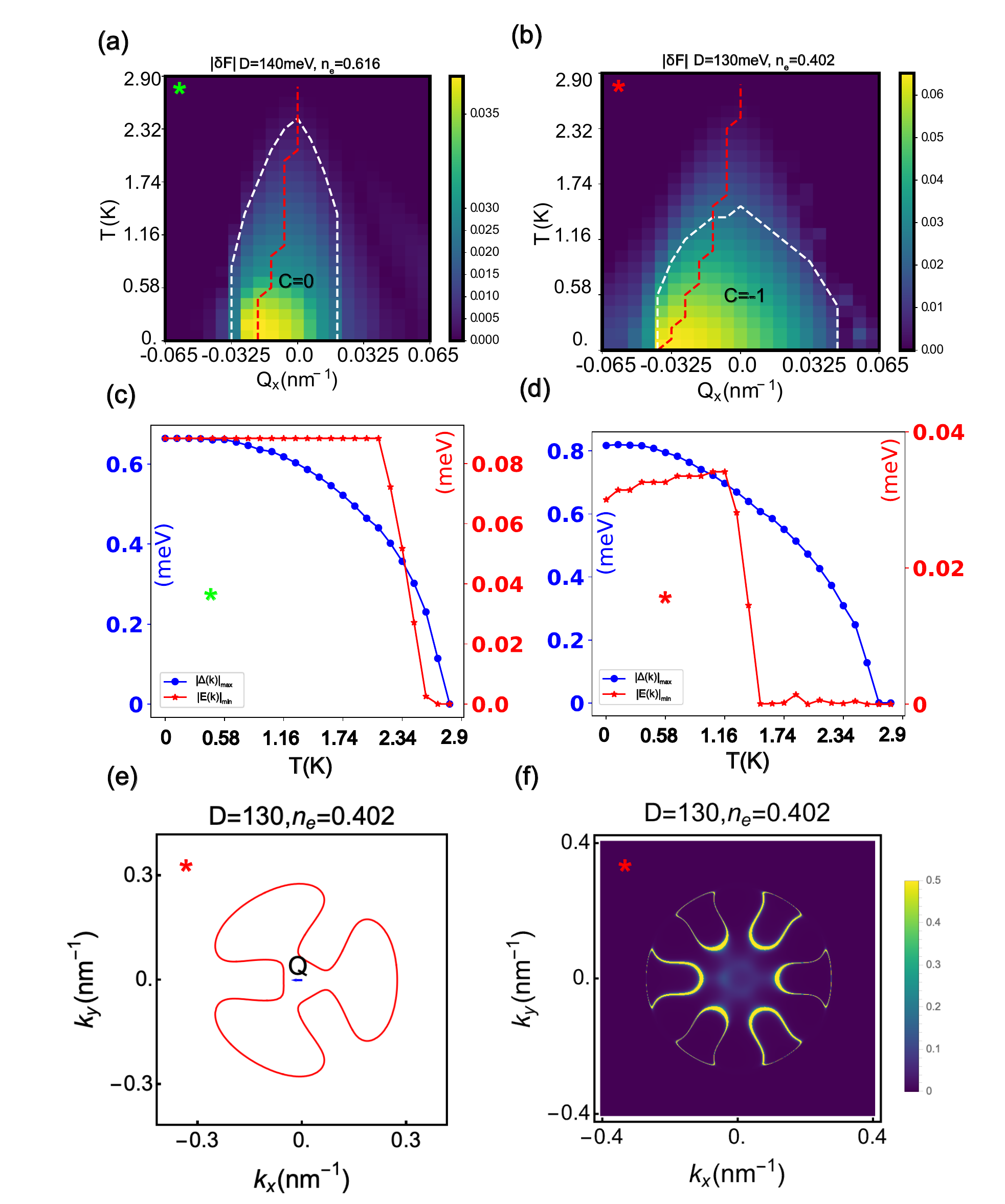}
    \caption{(a), (b) the free energy densities of $(D,n)=(140,0.616)$ and $(D,n)=(130,0.402)$ at $\epsilon=6$, respectively, corresponding to the Green and red stars in Fig.~\ref{fig:phase_diagram}(a). We have $\delta F\leq 0$ so $|\delta F|=-\delta F$. The red dashed lines correspond to the optimal $Q_x$ with lowest free energy density. (c) and (d) the temperature dependence of maximal order parameter $|\Delta(k)|_{\text{max}}$ (blue circles) and minimum quasi-particle gap $|E(\mathbf k)|_{\text{min}}$ (red stars) along the red dashed line in (a) and (b), respectively. (e) the normal state Fermi surface of $(D,n)=(130,0.402)$. The Blue arrow corresponds to the optimal ${\bf Q}$, which is about $Q_x=-0.03nm^{-1}$ at zero temperature (f) the spectral function $A(\omega=0,{\bf k})$ of $(D,n)=(130,0.402)$ at $T=2.33K$ and optimal $Q_x=-0.013nm^{-1}$. We can see the Bogoliubov Fermi surface.}
    \label{fig:finite_Q}
\end{figure}

\section{Incommensurate FFLO with $\mathbf Q\neq 0$}

So far we have restricted to ansatz with $\mathbf Q=0$. Because we define the momentum $\mathbf k$ relative to the valley $K$, even $\mathbf Q=0$ ansatz can be called a FFLO superconductor. But for the commensurate momentum, the $C_3$ symmetry is not broken and the translation symmetry is preserved combined with a gauge transformation. This is very different from the well-known FFLO superconductor with incommensurate momentum.  Here we show that actually the incommensurate FFLO with $\mathbf Q\neq 0$ is the ground state.  Because our model only has $C_3$ rotation symmetry and $\epsilon(\mathbf k)\neq \epsilon(-\mathbf k)$, there is no reason to assume $\mathbf Q=0$ as the optimal momentum of the Cooper pair.

In order to study the finite momentum pairing, we solve the self-consistent equation Eq.~\ref{eq:self_consistent} for different momentum ${\bf Q}$ and select the one with the lowest free energy. The free energy is defined as
\begin{align}
    F_\mathbf Q(\Delta_\mathbf Q({\bf k}))=&-T\sum_{k>0}\log(1+e^{-\beta(e_0\pm \eta_k)})+\sum_{k>0}(\epsilon_{-k+Q/2}-\mu)\nonumber\\
    -&\frac{1}{2A}\sum_{k,k^\prime}\tilde{V}(k^\prime-k)\langle c^\dagger_{k_1+\frac{Q}{2}}c^\dagger_{-k_1+\frac{Q}{2}}\rangle\langle c_{-k_2+\frac{Q}{2}}c_{k_2+\frac{Q}{2}}\rangle,
\end{align}
where the second line is the condensation energy. We calculate the free energy difference between the superconducting state and the normal state $\delta F_\mathbf Q=F_\mathbf Q(\Delta({\bf k}))-F(0)$ as $F(0)$ does not depend on $\mathbf Q$. We plot $\delta F_\mathbf Q$ in the parameter space of $(Q_x, T)$ by fixing $Q_y=0$ for two representative $(n,D)$ parameters labeled as red and blue stars in Fig.~\ref{fig:phase_diagram}(c).  As shown in Fig.~\ref{fig:finite_Q}(a) and Fig.~\ref{fig:finite_Q}(b), we find that optimal $Q_x$ is small but non-zero at low temperature.  $Q_x$ is at order $0.03\ nm^{-1}$ and can reach $14\%$ of the maximal $k_F$ for $(D,n_e)=(130,0.402)$\footnote{The maximal $k_F$ is defined as the largest $|\mathbf k|$ on the Fermi surface.}.

We confirmed that the ansatz with a small finite $\mathbf Q$ has roughly the same Bogoliubov quasiparticle spectrum as the $\mathbf Q=0$ ansatz. Especially there is still a $C=-1$ to $C=0$ topological phase transition. Therefore we have a topological incommensurate FFLO superconductor with $C=-1$.

We show the temperature dependence of $|\Delta(\mathbf k)|_{\text{max}}$ and the Bogoliubov quasiparticle gap for the $C=0$ and $C=-1$ ansatz with optimal $\mathbf Q$ in Fig.~\ref{fig:finite_Q}(c) and (d), respectively. The optimal $|\mathbf Q|$ gradually decreases to zero with temperature.  The maximal $|\Delta(\mathbf k)|$ also decreases with temperature and vanishes at the mean field $T_c$, which is actually how we decided the value of $T_c$ for Fig.~\ref{fig:phase_diagram}(d). However, for the $C=-1$ ansatz in Fig.~\ref{fig:finite_Q}(b)(d), we notice that the quasi-particle gap closes well before $T_c$, implying a gapless superconductor at higher temperature with Bogoliubov Fermi surface. We show the normal state Fermi surface in Fig.~\ref{fig:finite_Q}(e) and the spectral weight of the Bogoliubov Fermi surface at a temperature below $T_c$ in Fig.~\ref{fig:finite_Q}(f). Clearly the low temperature topological superconductor transits to a gapless superconductor. Hence the tetra-layer graphene offers an opportunity to study both the topological superconductor and the Bogoliubov Fermi surfaces, coexisting with incommensurate FFLO order parameter.

\section{Conclusion}
Through RPA calculation, we demonstrated the emergence of chiral superconductivity in spin-valley-polarized tetra-layer rhombohedral graphene via the Kohn-Luttinger mechanism. By varying the displacement field and electron density, we mapped out the phase diagram and identified a topological phase transition from a chiral superconducting phase with $C=-1$ to a trivial one with $C=0$. Besides, a gapless superconductor with Bogoliubov Fermi surfaces shows up at higher temperature just below $T_c$. Meanwhile we show that the optimal Cooper pair momentum $\mathbf Q$ is small but finite, indicating an incommensurate FFLO order parameter. We have restricted to the FF ansatz with a single $\mathbf Q$, so the superconductor has a nematic order, which may be detected by the anisotropy of the critical current along different directions. We leave to future to decide whether a $C_3$ symmetric LO phase with a CDW order is favored or not. We conclude that the system has a richer physics than a simple fully gapped p-wave topological superconductor. From a theoretical perspective, the RPA calculation is only a crude approximation and can not capture other competing states such as a Wigner crystal on equal footing. We leave it to future to develop an unbiased framework to incorporate both chiral superconductor and topological or trivial Wigner crystals.

\section{Acknowledgement}
This work was supported by a
startup fund from Johns Hopkins University and the
Alfred P. Sloan Foundation through a Sloan Research
Fellowship (YHZ).

\bibliographystyle{apsrev4-1}
\bibliography{ref}

\begin{thebibliography}{83}%
\makeatletter
\providecommand \@ifxundefined [1]{%
 \@ifx{#1\undefined}
}%
\providecommand \@ifnum [1]{%
 \ifnum #1\expandafter \@firstoftwo
 \else \expandafter \@secondoftwo
 \fi
}%
\providecommand \@ifx [1]{%
 \ifx #1\expandafter \@firstoftwo
 \else \expandafter \@secondoftwo
 \fi
}%
\providecommand \natexlab [1]{#1}%
\providecommand \enquote  [1]{``#1''}%
\providecommand \bibnamefont  [1]{#1}%
\providecommand \bibfnamefont [1]{#1}%
\providecommand \citenamefont [1]{#1}%
\providecommand \href@noop [0]{\@secondoftwo}%
\providecommand \href [0]{\begingroup \@sanitize@url \@href}%
\providecommand \@href[1]{\@@startlink{#1}\@@href}%
\providecommand \@@href[1]{\endgroup#1\@@endlink}%
\providecommand \@sanitize@url [0]{\catcode `\\12\catcode `\$12\catcode
  `\&12\catcode `\#12\catcode `\^12\catcode `\_12\catcode `\%12\relax}%
\providecommand \@@startlink[1]{}%
\providecommand \@@endlink[0]{}%
\providecommand \url  [0]{\begingroup\@sanitize@url \@url }%
\providecommand \@url [1]{\endgroup\@href {#1}{\urlprefix }}%
\providecommand \urlprefix  [0]{URL }%
\providecommand \Eprint [0]{\href }%
\providecommand \doibase [0]{http://dx.doi.org/}%
\providecommand \selectlanguage [0]{\@gobble}%
\providecommand \bibinfo  [0]{\@secondoftwo}%
\providecommand \bibfield  [0]{\@secondoftwo}%
\providecommand \translation [1]{[#1]}%
\providecommand \BibitemOpen [0]{}%
\providecommand \bibitemStop [0]{}%
\providecommand \bibitemNoStop [0]{.\EOS\space}%
\providecommand \EOS [0]{\spacefactor3000\relax}%
\providecommand \BibitemShut  [1]{\csname bibitem#1\endcsname}%
\let\auto@bib@innerbib\@empty
\bibitem [{\citenamefont {Sigrist}\ and\ \citenamefont
  {Ueda}(1991)}]{RevModPhys.63.239}%
  \BibitemOpen
  \bibfield  {author} {\bibinfo {author} {\bibfnamefont {M.}~\bibnamefont
  {Sigrist}}\ and\ \bibinfo {author} {\bibfnamefont {K.}~\bibnamefont {Ueda}},\
  }\href {\doibase 10.1103/RevModPhys.63.239} {\bibfield  {journal} {\bibinfo
  {journal} {Rev. Mod. Phys.}\ }\textbf {\bibinfo {volume} {63}},\ \bibinfo
  {pages} {239} (\bibinfo {year} {1991})}\BibitemShut {NoStop}%
\bibitem [{\citenamefont {Alicea}(2012)}]{Alicea_2012}%
  \BibitemOpen
  \bibfield  {author} {\bibinfo {author} {\bibfnamefont {J.}~\bibnamefont
  {Alicea}},\ }\href {\doibase 10.1088/0034-4885/75/7/076501} {\bibfield
  {journal} {\bibinfo  {journal} {Reports on Progress in Physics}\ }\textbf
  {\bibinfo {volume} {75}},\ \bibinfo {pages} {076501} (\bibinfo {year}
  {2012})}\BibitemShut {NoStop}%
\bibitem [{\citenamefont {Nayak}\ \emph {et~al.}(2008)\citenamefont {Nayak},
  \citenamefont {Simon}, \citenamefont {Stern}, \citenamefont {Freedman},\ and\
  \citenamefont {Das~Sarma}}]{Nayak2008Non}%
  \BibitemOpen
  \bibfield  {author} {\bibinfo {author} {\bibfnamefont {C.}~\bibnamefont
  {Nayak}}, \bibinfo {author} {\bibfnamefont {S.~H.}\ \bibnamefont {Simon}},
  \bibinfo {author} {\bibfnamefont {A.}~\bibnamefont {Stern}}, \bibinfo
  {author} {\bibfnamefont {M.}~\bibnamefont {Freedman}}, \ and\ \bibinfo
  {author} {\bibfnamefont {S.}~\bibnamefont {Das~Sarma}},\ }\href@noop {}
  {\bibfield  {journal} {\bibinfo  {journal} {Reviews of Modern Physics}\
  }\textbf {\bibinfo {volume} {80}},\ \bibinfo {pages} {1083} (\bibinfo {year}
  {2008})}\BibitemShut {NoStop}%
\bibitem [{\citenamefont {Qi}\ and\ \citenamefont {Zhang}(2011)}]{ZhangRMP}%
  \BibitemOpen
  \bibfield  {author} {\bibinfo {author} {\bibfnamefont {X.-L.}\ \bibnamefont
  {Qi}}\ and\ \bibinfo {author} {\bibfnamefont {S.-C.}\ \bibnamefont {Zhang}},\
  }\href {\doibase 10.1103/RevModPhys.83.1057} {\bibfield  {journal} {\bibinfo
  {journal} {Rev. Mod. Phys.}\ }\textbf {\bibinfo {volume} {83}},\ \bibinfo
  {pages} {1057} (\bibinfo {year} {2011})}\BibitemShut {NoStop}%
\bibitem [{\citenamefont {Hasan}\ and\ \citenamefont
  {Kane}(2010)}]{RevModPhys.82.3045}%
  \BibitemOpen
  \bibfield  {author} {\bibinfo {author} {\bibfnamefont {M.~Z.}\ \bibnamefont
  {Hasan}}\ and\ \bibinfo {author} {\bibfnamefont {C.~L.}\ \bibnamefont
  {Kane}},\ }\href {\doibase 10.1103/RevModPhys.82.3045} {\bibfield  {journal}
  {\bibinfo  {journal} {Rev. Mod. Phys.}\ }\textbf {\bibinfo {volume} {82}},\
  \bibinfo {pages} {3045} (\bibinfo {year} {2010})}\BibitemShut {NoStop}%
\bibitem [{\citenamefont {Read}\ and\ \citenamefont
  {Green}(2000)}]{Read2000paired}%
  \BibitemOpen
  \bibfield  {author} {\bibinfo {author} {\bibfnamefont {N.}~\bibnamefont
  {Read}}\ and\ \bibinfo {author} {\bibfnamefont {D.}~\bibnamefont {Green}},\
  }\href@noop {} {\bibfield  {journal} {\bibinfo  {journal} {Physical Review
  B}\ }\textbf {\bibinfo {volume} {61}},\ \bibinfo {pages} {10267} (\bibinfo
  {year} {2000})}\BibitemShut {NoStop}%
\bibitem [{\citenamefont {Schnyder}\ \emph {et~al.}(2008)\citenamefont
  {Schnyder}, \citenamefont {Ryu}, \citenamefont {Furusaki},\ and\
  \citenamefont {Ludwig}}]{Schnyder2008Classification}%
  \BibitemOpen
  \bibfield  {author} {\bibinfo {author} {\bibfnamefont {A.~P.}\ \bibnamefont
  {Schnyder}}, \bibinfo {author} {\bibfnamefont {S.}~\bibnamefont {Ryu}},
  \bibinfo {author} {\bibfnamefont {A.}~\bibnamefont {Furusaki}}, \ and\
  \bibinfo {author} {\bibfnamefont {A.~W.~W.}\ \bibnamefont {Ludwig}},\ }\href
  {\doibase 10.1103/PhysRevB.78.195125} {\bibfield  {journal} {\bibinfo
  {journal} {Phys. Rev. B}\ }\textbf {\bibinfo {volume} {78}},\ \bibinfo
  {pages} {195125} (\bibinfo {year} {2008})}\BibitemShut {NoStop}%
\bibitem [{\citenamefont {Kitaev}(2009)}]{Kitaev2009Periodic}%
  \BibitemOpen
  \bibfield  {author} {\bibinfo {author} {\bibfnamefont {A.}~\bibnamefont
  {Kitaev}},\ }in\ \href@noop {} {\emph {\bibinfo {booktitle} {AIP conference
  proceedings}}},\ Vol.\ \bibinfo {volume} {1134}\ (\bibinfo {organization}
  {American Institute of Physics},\ \bibinfo {year} {2009})\ pp.\ \bibinfo
  {pages} {22--30}\BibitemShut {NoStop}%
\bibitem [{\citenamefont {Sato}\ and\ \citenamefont
  {Ando}(2017)}]{Sato2017topological}%
  \BibitemOpen
  \bibfield  {author} {\bibinfo {author} {\bibfnamefont {M.}~\bibnamefont
  {Sato}}\ and\ \bibinfo {author} {\bibfnamefont {Y.}~\bibnamefont {Ando}},\
  }\href@noop {} {\bibfield  {journal} {\bibinfo  {journal} {Reports on
  Progress in Physics}\ }\textbf {\bibinfo {volume} {80}},\ \bibinfo {pages}
  {076501} (\bibinfo {year} {2017})}\BibitemShut {NoStop}%
\bibitem [{\citenamefont {Maeno}\ \emph {et~al.}(1994)\citenamefont {Maeno},
  \citenamefont {Hashimoto}, \citenamefont {Yoshida}, \citenamefont
  {Nishizaki}, \citenamefont {Fujita}, \citenamefont {Bednorz},\ and\
  \citenamefont {Lichtenberg}}]{Maeno1994}%
  \BibitemOpen
  \bibfield  {author} {\bibinfo {author} {\bibfnamefont {Y.}~\bibnamefont
  {Maeno}}, \bibinfo {author} {\bibfnamefont {H.}~\bibnamefont {Hashimoto}},
  \bibinfo {author} {\bibfnamefont {K.}~\bibnamefont {Yoshida}}, \bibinfo
  {author} {\bibfnamefont {S.}~\bibnamefont {Nishizaki}}, \bibinfo {author}
  {\bibfnamefont {T.}~\bibnamefont {Fujita}}, \bibinfo {author} {\bibfnamefont
  {J.~G.}\ \bibnamefont {Bednorz}}, \ and\ \bibinfo {author} {\bibfnamefont
  {F.}~\bibnamefont {Lichtenberg}},\ }\href {\doibase 10.1038/372532a0}
  {\bibfield  {journal} {\bibinfo  {journal} {Nature}\ }\textbf {\bibinfo
  {volume} {372}},\ \bibinfo {pages} {532} (\bibinfo {year}
  {1994})}\BibitemShut {NoStop}%
\bibitem [{\citenamefont {de~Visser}(2019)}]{doi:10.7566/JPSJNC.16.08}%
  \BibitemOpen
  \bibfield  {author} {\bibinfo {author} {\bibfnamefont {A.}~\bibnamefont
  {de~Visser}},\ }\href {\doibase 10.7566/JPSJNC.16.08} {\bibfield  {journal}
  {\bibinfo  {journal} {JPSJ News and Comments}\ }\textbf {\bibinfo {volume}
  {16}},\ \bibinfo {pages} {08} (\bibinfo {year} {2019})},\ \Eprint
  {http://arxiv.org/abs/https://doi.org/10.7566/JPSJNC.16.08}
  {https://doi.org/10.7566/JPSJNC.16.08} \BibitemShut {NoStop}%
\bibitem [{\citenamefont {Jiao}\ \emph {et~al.}(2020)\citenamefont {Jiao},
  \citenamefont {Howard}, \citenamefont {Ran}, \citenamefont {Wang},
  \citenamefont {Rodriguez}, \citenamefont {Sigrist}, \citenamefont {Wang},
  \citenamefont {Butch},\ and\ \citenamefont {Madhavan}}]{Jiao2020}%
  \BibitemOpen
  \bibfield  {author} {\bibinfo {author} {\bibfnamefont {L.}~\bibnamefont
  {Jiao}}, \bibinfo {author} {\bibfnamefont {S.}~\bibnamefont {Howard}},
  \bibinfo {author} {\bibfnamefont {S.}~\bibnamefont {Ran}}, \bibinfo {author}
  {\bibfnamefont {Z.}~\bibnamefont {Wang}}, \bibinfo {author} {\bibfnamefont
  {J.~O.}\ \bibnamefont {Rodriguez}}, \bibinfo {author} {\bibfnamefont
  {M.}~\bibnamefont {Sigrist}}, \bibinfo {author} {\bibfnamefont
  {Z.}~\bibnamefont {Wang}}, \bibinfo {author} {\bibfnamefont {N.~P.}\
  \bibnamefont {Butch}}, \ and\ \bibinfo {author} {\bibfnamefont
  {V.}~\bibnamefont {Madhavan}},\ }\href {\doibase 10.1038/s41586-020-2122-2}
  {\bibfield  {journal} {\bibinfo  {journal} {Nature}\ }\textbf {\bibinfo
  {volume} {579}},\ \bibinfo {pages} {523} (\bibinfo {year}
  {2020})}\BibitemShut {NoStop}%
\bibitem [{\citenamefont {Aoki}\ \emph {et~al.}(2022)\citenamefont {Aoki},
  \citenamefont {Brison}, \citenamefont {Flouquet}, \citenamefont {Ishida},
  \citenamefont {Knebel}, \citenamefont {Tokunaga},\ and\ \citenamefont
  {Yanase}}]{Aoki_2022}%
  \BibitemOpen
  \bibfield  {author} {\bibinfo {author} {\bibfnamefont {D.}~\bibnamefont
  {Aoki}}, \bibinfo {author} {\bibfnamefont {J.-P.}\ \bibnamefont {Brison}},
  \bibinfo {author} {\bibfnamefont {J.}~\bibnamefont {Flouquet}}, \bibinfo
  {author} {\bibfnamefont {K.}~\bibnamefont {Ishida}}, \bibinfo {author}
  {\bibfnamefont {G.}~\bibnamefont {Knebel}}, \bibinfo {author} {\bibfnamefont
  {Y.}~\bibnamefont {Tokunaga}}, \ and\ \bibinfo {author} {\bibfnamefont
  {Y.}~\bibnamefont {Yanase}},\ }\href {\doibase 10.1088/1361-648X/ac5863}
  {\bibfield  {journal} {\bibinfo  {journal} {Journal of Physics: Condensed
  Matter}\ }\textbf {\bibinfo {volume} {34}},\ \bibinfo {pages} {243002}
  (\bibinfo {year} {2022})}\BibitemShut {NoStop}%
\bibitem [{\citenamefont {Theuss}\ \emph {et~al.}(2024)\citenamefont {Theuss},
  \citenamefont {Shragai}, \citenamefont {Grissonnanche}, \citenamefont
  {Hayes}, \citenamefont {Saha}, \citenamefont {Eo}, \citenamefont {Suarez},
  \citenamefont {Shishidou}, \citenamefont {Butch}, \citenamefont {Paglione},\
  and\ \citenamefont {Ramshaw}}]{Theuss2024}%
  \BibitemOpen
  \bibfield  {author} {\bibinfo {author} {\bibfnamefont {F.}~\bibnamefont
  {Theuss}}, \bibinfo {author} {\bibfnamefont {A.}~\bibnamefont {Shragai}},
  \bibinfo {author} {\bibfnamefont {G.}~\bibnamefont {Grissonnanche}}, \bibinfo
  {author} {\bibfnamefont {I.~M.}\ \bibnamefont {Hayes}}, \bibinfo {author}
  {\bibfnamefont {S.~R.}\ \bibnamefont {Saha}}, \bibinfo {author}
  {\bibfnamefont {Y.~S.}\ \bibnamefont {Eo}}, \bibinfo {author} {\bibfnamefont
  {A.}~\bibnamefont {Suarez}}, \bibinfo {author} {\bibfnamefont
  {T.}~\bibnamefont {Shishidou}}, \bibinfo {author} {\bibfnamefont {N.~P.}\
  \bibnamefont {Butch}}, \bibinfo {author} {\bibfnamefont {J.}~\bibnamefont
  {Paglione}}, \ and\ \bibinfo {author} {\bibfnamefont {B.~J.}\ \bibnamefont
  {Ramshaw}},\ }\href {\doibase 10.1038/s41567-024-02493-1} {\bibfield
  {journal} {\bibinfo  {journal} {Nature Physics}\ }\textbf {\bibinfo {volume}
  {20}},\ \bibinfo {pages} {1124} (\bibinfo {year} {2024})}\BibitemShut
  {NoStop}%
\bibitem [{\citenamefont {Li}\ \emph {et~al.}(2021)\citenamefont {Li},
  \citenamefont {Kikugawa}, \citenamefont {Sokolov}, \citenamefont
  {Jerzembeck}, \citenamefont {Gibbs}, \citenamefont {Maeno}, \citenamefont
  {Hicks}, \citenamefont {Schmalian}, \citenamefont {Nicklas},\ and\
  \citenamefont {Mackenzie}}]{doi:10.1073/pnas.2020492118}%
  \BibitemOpen
  \bibfield  {author} {\bibinfo {author} {\bibfnamefont {Y.-S.}\ \bibnamefont
  {Li}}, \bibinfo {author} {\bibfnamefont {N.}~\bibnamefont {Kikugawa}},
  \bibinfo {author} {\bibfnamefont {D.~A.}\ \bibnamefont {Sokolov}}, \bibinfo
  {author} {\bibfnamefont {F.}~\bibnamefont {Jerzembeck}}, \bibinfo {author}
  {\bibfnamefont {A.~S.}\ \bibnamefont {Gibbs}}, \bibinfo {author}
  {\bibfnamefont {Y.}~\bibnamefont {Maeno}}, \bibinfo {author} {\bibfnamefont
  {C.~W.}\ \bibnamefont {Hicks}}, \bibinfo {author} {\bibfnamefont
  {J.}~\bibnamefont {Schmalian}}, \bibinfo {author} {\bibfnamefont
  {M.}~\bibnamefont {Nicklas}}, \ and\ \bibinfo {author} {\bibfnamefont
  {A.~P.}\ \bibnamefont {Mackenzie}},\ }\href {\doibase
  10.1073/pnas.2020492118} {\bibfield  {journal} {\bibinfo  {journal}
  {Proceedings of the National Academy of Sciences}\ }\textbf {\bibinfo
  {volume} {118}},\ \bibinfo {pages} {e2020492118} (\bibinfo {year} {2021})},\
  \Eprint
  {http://arxiv.org/abs/https://www.pnas.org/doi/pdf/10.1073/pnas.2020492118}
  {https://www.pnas.org/doi/pdf/10.1073/pnas.2020492118} \BibitemShut {NoStop}%
\bibitem [{\citenamefont {Pustogow}\ \emph {et~al.}(2019)\citenamefont
  {Pustogow}, \citenamefont {Luo}, \citenamefont {Chronister}, \citenamefont
  {Su}, \citenamefont {Sokolov}, \citenamefont {Jerzembeck}, \citenamefont
  {Mackenzie}, \citenamefont {Hicks}, \citenamefont {Kikugawa}, \citenamefont
  {Raghu}, \citenamefont {Bauer},\ and\ \citenamefont {Brown}}]{Pustogow2019}%
  \BibitemOpen
  \bibfield  {author} {\bibinfo {author} {\bibfnamefont {A.}~\bibnamefont
  {Pustogow}}, \bibinfo {author} {\bibfnamefont {Y.}~\bibnamefont {Luo}},
  \bibinfo {author} {\bibfnamefont {A.}~\bibnamefont {Chronister}}, \bibinfo
  {author} {\bibfnamefont {Y.-S.}\ \bibnamefont {Su}}, \bibinfo {author}
  {\bibfnamefont {D.~A.}\ \bibnamefont {Sokolov}}, \bibinfo {author}
  {\bibfnamefont {F.}~\bibnamefont {Jerzembeck}}, \bibinfo {author}
  {\bibfnamefont {A.~P.}\ \bibnamefont {Mackenzie}}, \bibinfo {author}
  {\bibfnamefont {C.~W.}\ \bibnamefont {Hicks}}, \bibinfo {author}
  {\bibfnamefont {N.}~\bibnamefont {Kikugawa}}, \bibinfo {author}
  {\bibfnamefont {S.}~\bibnamefont {Raghu}}, \bibinfo {author} {\bibfnamefont
  {E.~D.}\ \bibnamefont {Bauer}}, \ and\ \bibinfo {author} {\bibfnamefont
  {S.~E.}\ \bibnamefont {Brown}},\ }\href {\doibase 10.1038/s41586-019-1596-2}
  {\bibfield  {journal} {\bibinfo  {journal} {Nature}\ }\textbf {\bibinfo
  {volume} {574}},\ \bibinfo {pages} {72} (\bibinfo {year} {2019})}\BibitemShut
  {NoStop}%
\bibitem [{\citenamefont {Willa}\ \emph {et~al.}(2021)\citenamefont {Willa},
  \citenamefont {Hecker}, \citenamefont {Fernandes},\ and\ \citenamefont
  {Schmalian}}]{PhysRevB.104.024511}%
  \BibitemOpen
  \bibfield  {author} {\bibinfo {author} {\bibfnamefont {R.}~\bibnamefont
  {Willa}}, \bibinfo {author} {\bibfnamefont {M.}~\bibnamefont {Hecker}},
  \bibinfo {author} {\bibfnamefont {R.~M.}\ \bibnamefont {Fernandes}}, \ and\
  \bibinfo {author} {\bibfnamefont {J.}~\bibnamefont {Schmalian}},\ }\href
  {\doibase 10.1103/PhysRevB.104.024511} {\bibfield  {journal} {\bibinfo
  {journal} {Phys. Rev. B}\ }\textbf {\bibinfo {volume} {104}},\ \bibinfo
  {pages} {024511} (\bibinfo {year} {2021})}\BibitemShut {NoStop}%
\bibitem [{\citenamefont {Andrei}\ \emph {et~al.}(2021)\citenamefont {Andrei},
  \citenamefont {Efetov}, \citenamefont {Jarillo-Herrero}, \citenamefont
  {MacDonald}, \citenamefont {Mak}, \citenamefont {Senthil}, \citenamefont
  {Tutuc}, \citenamefont {Yazdani},\ and\ \citenamefont {Young}}]{Andrei2021}%
  \BibitemOpen
  \bibfield  {author} {\bibinfo {author} {\bibfnamefont {E.~Y.}\ \bibnamefont
  {Andrei}}, \bibinfo {author} {\bibfnamefont {D.~K.}\ \bibnamefont {Efetov}},
  \bibinfo {author} {\bibfnamefont {P.}~\bibnamefont {Jarillo-Herrero}},
  \bibinfo {author} {\bibfnamefont {A.~H.}\ \bibnamefont {MacDonald}}, \bibinfo
  {author} {\bibfnamefont {K.~F.}\ \bibnamefont {Mak}}, \bibinfo {author}
  {\bibfnamefont {T.}~\bibnamefont {Senthil}}, \bibinfo {author} {\bibfnamefont
  {E.}~\bibnamefont {Tutuc}}, \bibinfo {author} {\bibfnamefont
  {A.}~\bibnamefont {Yazdani}}, \ and\ \bibinfo {author} {\bibfnamefont
  {A.~F.}\ \bibnamefont {Young}},\ }\href {\doibase 10.1038/s41578-021-00284-1}
  {\bibfield  {journal} {\bibinfo  {journal} {Nature Reviews Materials}\
  }\textbf {\bibinfo {volume} {6}},\ \bibinfo {pages} {201} (\bibinfo {year}
  {2021})}\BibitemShut {NoStop}%
\bibitem [{\citenamefont {Nuckolls}\ and\ \citenamefont
  {Yazdani}(2024)}]{Nuckolls2024}%
  \BibitemOpen
  \bibfield  {author} {\bibinfo {author} {\bibfnamefont {K.~P.}\ \bibnamefont
  {Nuckolls}}\ and\ \bibinfo {author} {\bibfnamefont {A.}~\bibnamefont
  {Yazdani}},\ }\href {\doibase 10.1038/s41578-024-00682-1} {\bibfield
  {journal} {\bibinfo  {journal} {Nature Reviews Materials}\ }\textbf {\bibinfo
  {volume} {9}},\ \bibinfo {pages} {460} (\bibinfo {year} {2024})}\BibitemShut
  {NoStop}%
\bibitem [{\citenamefont {Balents}\ \emph {et~al.}(2020)\citenamefont
  {Balents}, \citenamefont {Dean}, \citenamefont {Efetov},\ and\ \citenamefont
  {Young}}]{Balents2020}%
  \BibitemOpen
  \bibfield  {author} {\bibinfo {author} {\bibfnamefont {L.}~\bibnamefont
  {Balents}}, \bibinfo {author} {\bibfnamefont {C.~R.}\ \bibnamefont {Dean}},
  \bibinfo {author} {\bibfnamefont {D.~K.}\ \bibnamefont {Efetov}}, \ and\
  \bibinfo {author} {\bibfnamefont {A.~F.}\ \bibnamefont {Young}},\ }\href
  {\doibase 10.1038/s41567-020-0906-9} {\bibfield  {journal} {\bibinfo
  {journal} {Nature Physics}\ }\textbf {\bibinfo {volume} {16}},\ \bibinfo
  {pages} {725} (\bibinfo {year} {2020})}\BibitemShut {NoStop}%
\bibitem [{\citenamefont {Cao}\ \emph {et~al.}(2018)\citenamefont {Cao},
  \citenamefont {Fatemi}, \citenamefont {Fang}, \citenamefont {Watanabe},
  \citenamefont {Taniguchi}, \citenamefont {Kaxiras},\ and\ \citenamefont
  {Jarillo-Herrero}}]{Cao2018Unconventional}%
  \BibitemOpen
  \bibfield  {author} {\bibinfo {author} {\bibfnamefont {Y.}~\bibnamefont
  {Cao}}, \bibinfo {author} {\bibfnamefont {V.}~\bibnamefont {Fatemi}},
  \bibinfo {author} {\bibfnamefont {S.}~\bibnamefont {Fang}}, \bibinfo {author}
  {\bibfnamefont {K.}~\bibnamefont {Watanabe}}, \bibinfo {author}
  {\bibfnamefont {T.}~\bibnamefont {Taniguchi}}, \bibinfo {author}
  {\bibfnamefont {E.}~\bibnamefont {Kaxiras}}, \ and\ \bibinfo {author}
  {\bibfnamefont {P.}~\bibnamefont {Jarillo-Herrero}},\ }\href {\doibase
  10.1038/nature26160} {\bibfield  {journal} {\bibinfo  {journal} {Nature}\
  }\textbf {\bibinfo {volume} {556}},\ \bibinfo {pages} {43} (\bibinfo {year}
  {2018})}\BibitemShut {NoStop}%
\bibitem [{\citenamefont {Yankowitz}\ \emph
  {et~al.}(2019{\natexlab{a}})\citenamefont {Yankowitz}, \citenamefont {Chen},
  \citenamefont {Polshyn}, \citenamefont {Zhang}, \citenamefont {Watanabe},
  \citenamefont {Taniguchi}, \citenamefont {Graf}, \citenamefont {Young},\ and\
  \citenamefont {Dean}}]{Yankowitz2019Tuning}%
  \BibitemOpen
  \bibfield  {author} {\bibinfo {author} {\bibfnamefont {M.}~\bibnamefont
  {Yankowitz}}, \bibinfo {author} {\bibfnamefont {S.}~\bibnamefont {Chen}},
  \bibinfo {author} {\bibfnamefont {H.}~\bibnamefont {Polshyn}}, \bibinfo
  {author} {\bibfnamefont {Y.}~\bibnamefont {Zhang}}, \bibinfo {author}
  {\bibfnamefont {K.}~\bibnamefont {Watanabe}}, \bibinfo {author}
  {\bibfnamefont {T.}~\bibnamefont {Taniguchi}}, \bibinfo {author}
  {\bibfnamefont {D.}~\bibnamefont {Graf}}, \bibinfo {author} {\bibfnamefont
  {A.~F.}\ \bibnamefont {Young}}, \ and\ \bibinfo {author} {\bibfnamefont
  {C.~R.}\ \bibnamefont {Dean}},\ }\href {\doibase 10.1126/science.aav1910}
  {\bibfield  {journal} {\bibinfo  {journal} {Science}\ }\textbf {\bibinfo
  {volume} {363}},\ \bibinfo {pages} {1059} (\bibinfo {year}
  {2019}{\natexlab{a}})}\BibitemShut {NoStop}%
\bibitem [{\citenamefont {Lu}\ \emph {et~al.}(2019)\citenamefont {Lu},
  \citenamefont {Stepanov}, \citenamefont {Yang}, \citenamefont {Xie},
  \citenamefont {Aamir}, \citenamefont {Das}, \citenamefont {Urgell},
  \citenamefont {Watanabe}, \citenamefont {Taniguchi}, \citenamefont {Zhang},
  \citenamefont {Bachtold}, \citenamefont {MacDonald},\ and\ \citenamefont
  {Efetov}}]{Lu2019Superconductors}%
  \BibitemOpen
  \bibfield  {author} {\bibinfo {author} {\bibfnamefont {X.}~\bibnamefont
  {Lu}}, \bibinfo {author} {\bibfnamefont {P.}~\bibnamefont {Stepanov}},
  \bibinfo {author} {\bibfnamefont {W.}~\bibnamefont {Yang}}, \bibinfo {author}
  {\bibfnamefont {M.}~\bibnamefont {Xie}}, \bibinfo {author} {\bibfnamefont
  {M.~A.}\ \bibnamefont {Aamir}}, \bibinfo {author} {\bibfnamefont
  {I.}~\bibnamefont {Das}}, \bibinfo {author} {\bibfnamefont {C.}~\bibnamefont
  {Urgell}}, \bibinfo {author} {\bibfnamefont {K.}~\bibnamefont {Watanabe}},
  \bibinfo {author} {\bibfnamefont {T.}~\bibnamefont {Taniguchi}}, \bibinfo
  {author} {\bibfnamefont {G.}~\bibnamefont {Zhang}}, \bibinfo {author}
  {\bibfnamefont {A.}~\bibnamefont {Bachtold}}, \bibinfo {author}
  {\bibfnamefont {A.~H.}\ \bibnamefont {MacDonald}}, \ and\ \bibinfo {author}
  {\bibfnamefont {D.~K.}\ \bibnamefont {Efetov}},\ }\href {\doibase
  10.1038/s41586-019-1695-0} {\bibfield  {journal} {\bibinfo  {journal}
  {Nature}\ }\textbf {\bibinfo {volume} {574}},\ \bibinfo {pages} {653}
  (\bibinfo {year} {2019})}\BibitemShut {NoStop}%
\bibitem [{\citenamefont {Han}\ \emph {et~al.}(2024)\citenamefont {Han},
  \citenamefont {Lu}, \citenamefont {Yao}, \citenamefont {Shi}, \citenamefont
  {Yang}, \citenamefont {Seo}, \citenamefont {Ye}, \citenamefont {Wu},
  \citenamefont {Zhou}, \citenamefont {Liu} \emph
  {et~al.}}]{Han2024signatures}%
  \BibitemOpen
  \bibfield  {author} {\bibinfo {author} {\bibfnamefont {T.}~\bibnamefont
  {Han}}, \bibinfo {author} {\bibfnamefont {Z.}~\bibnamefont {Lu}}, \bibinfo
  {author} {\bibfnamefont {Y.}~\bibnamefont {Yao}}, \bibinfo {author}
  {\bibfnamefont {L.}~\bibnamefont {Shi}}, \bibinfo {author} {\bibfnamefont
  {J.}~\bibnamefont {Yang}}, \bibinfo {author} {\bibfnamefont {J.}~\bibnamefont
  {Seo}}, \bibinfo {author} {\bibfnamefont {S.}~\bibnamefont {Ye}}, \bibinfo
  {author} {\bibfnamefont {Z.}~\bibnamefont {Wu}}, \bibinfo {author}
  {\bibfnamefont {M.}~\bibnamefont {Zhou}}, \bibinfo {author} {\bibfnamefont
  {H.}~\bibnamefont {Liu}},  \emph {et~al.},\ }\href@noop {} {\bibfield
  {journal} {\bibinfo  {journal} {arXiv preprint arXiv:2408.15233}\ } (\bibinfo
  {year} {2024})}\BibitemShut {NoStop}%
\bibitem [{\citenamefont {Choi}\ \emph {et~al.}(2024)\citenamefont {Choi},
  \citenamefont {Choi}, \citenamefont {Valentini}, \citenamefont {Patterson},
  \citenamefont {Holleis}, \citenamefont {Sheekey}, \citenamefont {Stoyanov},
  \citenamefont {Cheng}, \citenamefont {Taniguchi}, \citenamefont {Watanabe},\
  and\ \citenamefont {Young}}]{choi2024electric}%
  \BibitemOpen
  \bibfield  {author} {\bibinfo {author} {\bibfnamefont {Y.}~\bibnamefont
  {Choi}}, \bibinfo {author} {\bibfnamefont {Y.}~\bibnamefont {Choi}}, \bibinfo
  {author} {\bibfnamefont {M.}~\bibnamefont {Valentini}}, \bibinfo {author}
  {\bibfnamefont {C.~L.}\ \bibnamefont {Patterson}}, \bibinfo {author}
  {\bibfnamefont {L.~F.~W.}\ \bibnamefont {Holleis}}, \bibinfo {author}
  {\bibfnamefont {O.~I.}\ \bibnamefont {Sheekey}}, \bibinfo {author}
  {\bibfnamefont {H.}~\bibnamefont {Stoyanov}}, \bibinfo {author}
  {\bibfnamefont {X.}~\bibnamefont {Cheng}}, \bibinfo {author} {\bibfnamefont
  {T.}~\bibnamefont {Taniguchi}}, \bibinfo {author} {\bibfnamefont
  {K.}~\bibnamefont {Watanabe}}, \ and\ \bibinfo {author} {\bibfnamefont
  {A.~F.}\ \bibnamefont {Young}},\ }\href@noop {} {\enquote {\bibinfo {title}
  {Electric field control of superconductivity and quantized anomalous hall
  effects in rhombohedral tetralayer graphene},}\ } (\bibinfo {year} {2024}),\
  \Eprint {http://arxiv.org/abs/2408.12584} {arXiv:2408.12584
  [cond-mat.mes-hall]} \BibitemShut {NoStop}%
\bibitem [{\citenamefont {Xia}\ \emph {et~al.}(2024)\citenamefont {Xia},
  \citenamefont {Han}, \citenamefont {Watanabe}, \citenamefont {Taniguchi},
  \citenamefont {Shan},\ and\ \citenamefont {Mak}}]{Xia2024}%
  \BibitemOpen
  \bibfield  {author} {\bibinfo {author} {\bibfnamefont {Y.}~\bibnamefont
  {Xia}}, \bibinfo {author} {\bibfnamefont {Z.}~\bibnamefont {Han}}, \bibinfo
  {author} {\bibfnamefont {K.}~\bibnamefont {Watanabe}}, \bibinfo {author}
  {\bibfnamefont {T.}~\bibnamefont {Taniguchi}}, \bibinfo {author}
  {\bibfnamefont {J.}~\bibnamefont {Shan}}, \ and\ \bibinfo {author}
  {\bibfnamefont {K.~F.}\ \bibnamefont {Mak}},\ }\href {\doibase
  10.1038/s41586-024-08116-2} {\bibfield  {journal} {\bibinfo  {journal}
  {Nature}\ } (\bibinfo {year} {2024}),\
  10.1038/s41586-024-08116-2}\BibitemShut {NoStop}%
\bibitem [{\citenamefont {Guo}\ \emph {et~al.}(2024)\citenamefont {Guo},
  \citenamefont {Pack}, \citenamefont {Swann}, \citenamefont {Holtzman},
  \citenamefont {Cothrine}, \citenamefont {Watanabe}, \citenamefont
  {Taniguchi}, \citenamefont {Mandrus}, \citenamefont {Barmak}, \citenamefont
  {Hone} \emph {et~al.}}]{guo2024superconductivity}%
  \BibitemOpen
  \bibfield  {author} {\bibinfo {author} {\bibfnamefont {Y.}~\bibnamefont
  {Guo}}, \bibinfo {author} {\bibfnamefont {J.}~\bibnamefont {Pack}}, \bibinfo
  {author} {\bibfnamefont {J.}~\bibnamefont {Swann}}, \bibinfo {author}
  {\bibfnamefont {L.}~\bibnamefont {Holtzman}}, \bibinfo {author}
  {\bibfnamefont {M.}~\bibnamefont {Cothrine}}, \bibinfo {author}
  {\bibfnamefont {K.}~\bibnamefont {Watanabe}}, \bibinfo {author}
  {\bibfnamefont {T.}~\bibnamefont {Taniguchi}}, \bibinfo {author}
  {\bibfnamefont {D.}~\bibnamefont {Mandrus}}, \bibinfo {author} {\bibfnamefont
  {K.}~\bibnamefont {Barmak}}, \bibinfo {author} {\bibfnamefont
  {J.}~\bibnamefont {Hone}},  \emph {et~al.},\ }\href@noop {} {\bibfield
  {journal} {\bibinfo  {journal} {arXiv:2406.03418}\ } (\bibinfo {year}
  {2024})}\BibitemShut {NoStop}%
\bibitem [{\citenamefont {Jindal}\ \emph {et~al.}(2023)\citenamefont {Jindal},
  \citenamefont {Saha}, \citenamefont {Li}, \citenamefont {Taniguchi},
  \citenamefont {Watanabe}, \citenamefont {Hone}, \citenamefont {Birol},
  \citenamefont {Fernandes}, \citenamefont {Dean}, \citenamefont {Pasupathy},\
  and\ \citenamefont {Rhodes}}]{Jindal2023}%
  \BibitemOpen
  \bibfield  {author} {\bibinfo {author} {\bibfnamefont {A.}~\bibnamefont
  {Jindal}}, \bibinfo {author} {\bibfnamefont {A.}~\bibnamefont {Saha}},
  \bibinfo {author} {\bibfnamefont {Z.}~\bibnamefont {Li}}, \bibinfo {author}
  {\bibfnamefont {T.}~\bibnamefont {Taniguchi}}, \bibinfo {author}
  {\bibfnamefont {K.}~\bibnamefont {Watanabe}}, \bibinfo {author}
  {\bibfnamefont {J.~C.}\ \bibnamefont {Hone}}, \bibinfo {author}
  {\bibfnamefont {T.}~\bibnamefont {Birol}}, \bibinfo {author} {\bibfnamefont
  {R.~M.}\ \bibnamefont {Fernandes}}, \bibinfo {author} {\bibfnamefont {C.~R.}\
  \bibnamefont {Dean}}, \bibinfo {author} {\bibfnamefont {A.~N.}\ \bibnamefont
  {Pasupathy}}, \ and\ \bibinfo {author} {\bibfnamefont {D.~A.}\ \bibnamefont
  {Rhodes}},\ }\href {\doibase 10.1038/s41586-022-05521-3} {\bibfield
  {journal} {\bibinfo  {journal} {Nature}\ }\textbf {\bibinfo {volume} {613}},\
  \bibinfo {pages} {48} (\bibinfo {year} {2023})}\BibitemShut {NoStop}%
\bibitem [{\citenamefont {Jia}\ \emph {et~al.}(2024)\citenamefont {Jia},
  \citenamefont {Song}, \citenamefont {Zheng}, \citenamefont {Cheng},
  \citenamefont {Uzan}, \citenamefont {Yu}, \citenamefont {Tang}, \citenamefont
  {Pollak}, \citenamefont {Yuan}, \citenamefont {Onyszczak} \emph
  {et~al.}}]{jia2024anomalous}%
  \BibitemOpen
  \bibfield  {author} {\bibinfo {author} {\bibfnamefont {Y.}~\bibnamefont
  {Jia}}, \bibinfo {author} {\bibfnamefont {T.}~\bibnamefont {Song}}, \bibinfo
  {author} {\bibfnamefont {Z.~J.}\ \bibnamefont {Zheng}}, \bibinfo {author}
  {\bibfnamefont {G.}~\bibnamefont {Cheng}}, \bibinfo {author} {\bibfnamefont
  {A.~J.}\ \bibnamefont {Uzan}}, \bibinfo {author} {\bibfnamefont
  {G.}~\bibnamefont {Yu}}, \bibinfo {author} {\bibfnamefont {Y.}~\bibnamefont
  {Tang}}, \bibinfo {author} {\bibfnamefont {C.~J.}\ \bibnamefont {Pollak}},
  \bibinfo {author} {\bibfnamefont {F.}~\bibnamefont {Yuan}}, \bibinfo {author}
  {\bibfnamefont {M.}~\bibnamefont {Onyszczak}},  \emph {et~al.},\ }\href@noop
  {} {\bibfield  {journal} {\bibinfo  {journal} {arXiv:2409.04594}\ } (\bibinfo
  {year} {2024})}\BibitemShut {NoStop}%
\bibitem [{\citenamefont {Yankowitz}\ \emph
  {et~al.}(2019{\natexlab{b}})\citenamefont {Yankowitz}, \citenamefont {Chen},
  \citenamefont {Polshyn}, \citenamefont {Zhang}, \citenamefont {Watanabe},
  \citenamefont {Taniguchi}, \citenamefont {Graf}, \citenamefont {Young},\ and\
  \citenamefont {Dean}}]{doi:10.1126/science.aav1910}%
  \BibitemOpen
  \bibfield  {author} {\bibinfo {author} {\bibfnamefont {M.}~\bibnamefont
  {Yankowitz}}, \bibinfo {author} {\bibfnamefont {S.}~\bibnamefont {Chen}},
  \bibinfo {author} {\bibfnamefont {H.}~\bibnamefont {Polshyn}}, \bibinfo
  {author} {\bibfnamefont {Y.}~\bibnamefont {Zhang}}, \bibinfo {author}
  {\bibfnamefont {K.}~\bibnamefont {Watanabe}}, \bibinfo {author}
  {\bibfnamefont {T.}~\bibnamefont {Taniguchi}}, \bibinfo {author}
  {\bibfnamefont {D.}~\bibnamefont {Graf}}, \bibinfo {author} {\bibfnamefont
  {A.~F.}\ \bibnamefont {Young}}, \ and\ \bibinfo {author} {\bibfnamefont
  {C.~R.}\ \bibnamefont {Dean}},\ }\href {\doibase 10.1126/science.aav1910}
  {\bibfield  {journal} {\bibinfo  {journal} {Science}\ }\textbf {\bibinfo
  {volume} {363}},\ \bibinfo {pages} {1059} (\bibinfo {year}
  {2019}{\natexlab{b}})},\ \Eprint
  {http://arxiv.org/abs/https://www.science.org/doi/pdf/10.1126/science.aav1910}
  {https://www.science.org/doi/pdf/10.1126/science.aav1910} \BibitemShut
  {NoStop}%
\bibitem [{\citenamefont {Arora}\ \emph {et~al.}(2020)\citenamefont {Arora},
  \citenamefont {Polski}, \citenamefont {Zhang}, \citenamefont {Thomson},
  \citenamefont {Choi}, \citenamefont {Kim}, \citenamefont {Lin}, \citenamefont
  {Wilson}, \citenamefont {Xu}, \citenamefont {Chu}, \citenamefont {Watanabe},
  \citenamefont {Taniguchi}, \citenamefont {Alicea},\ and\ \citenamefont
  {Nadj-Perge}}]{Arora2020}%
  \BibitemOpen
  \bibfield  {author} {\bibinfo {author} {\bibfnamefont {H.~S.}\ \bibnamefont
  {Arora}}, \bibinfo {author} {\bibfnamefont {R.}~\bibnamefont {Polski}},
  \bibinfo {author} {\bibfnamefont {Y.}~\bibnamefont {Zhang}}, \bibinfo
  {author} {\bibfnamefont {A.}~\bibnamefont {Thomson}}, \bibinfo {author}
  {\bibfnamefont {Y.}~\bibnamefont {Choi}}, \bibinfo {author} {\bibfnamefont
  {H.}~\bibnamefont {Kim}}, \bibinfo {author} {\bibfnamefont {Z.}~\bibnamefont
  {Lin}}, \bibinfo {author} {\bibfnamefont {I.~Z.}\ \bibnamefont {Wilson}},
  \bibinfo {author} {\bibfnamefont {X.}~\bibnamefont {Xu}}, \bibinfo {author}
  {\bibfnamefont {J.-H.}\ \bibnamefont {Chu}}, \bibinfo {author} {\bibfnamefont
  {K.}~\bibnamefont {Watanabe}}, \bibinfo {author} {\bibfnamefont
  {T.}~\bibnamefont {Taniguchi}}, \bibinfo {author} {\bibfnamefont
  {J.}~\bibnamefont {Alicea}}, \ and\ \bibinfo {author} {\bibfnamefont
  {S.}~\bibnamefont {Nadj-Perge}},\ }\href {\doibase 10.1038/s41586-020-2473-8}
  {\bibfield  {journal} {\bibinfo  {journal} {Nature}\ }\textbf {\bibinfo
  {volume} {583}},\ \bibinfo {pages} {379} (\bibinfo {year}
  {2020})}\BibitemShut {NoStop}%
\bibitem [{\citenamefont {Cao}\ \emph {et~al.}(2021)\citenamefont {Cao},
  \citenamefont {Park}, \citenamefont {Watanabe}, \citenamefont {Taniguchi},\
  and\ \citenamefont {Jarillo-Herrero}}]{Cao2021}%
  \BibitemOpen
  \bibfield  {author} {\bibinfo {author} {\bibfnamefont {Y.}~\bibnamefont
  {Cao}}, \bibinfo {author} {\bibfnamefont {J.~M.}\ \bibnamefont {Park}},
  \bibinfo {author} {\bibfnamefont {K.}~\bibnamefont {Watanabe}}, \bibinfo
  {author} {\bibfnamefont {T.}~\bibnamefont {Taniguchi}}, \ and\ \bibinfo
  {author} {\bibfnamefont {P.}~\bibnamefont {Jarillo-Herrero}},\ }\href
  {\doibase 10.1038/s41586-021-03685-y} {\bibfield  {journal} {\bibinfo
  {journal} {Nature}\ }\textbf {\bibinfo {volume} {595}},\ \bibinfo {pages}
  {526} (\bibinfo {year} {2021})}\BibitemShut {NoStop}%
\bibitem [{\citenamefont {Oh}\ \emph {et~al.}(2021)\citenamefont {Oh},
  \citenamefont {Nuckolls}, \citenamefont {Wong}, \citenamefont {Lee},
  \citenamefont {Liu}, \citenamefont {Watanabe}, \citenamefont {Taniguchi},\
  and\ \citenamefont {Yazdani}}]{Oh2021}%
  \BibitemOpen
  \bibfield  {author} {\bibinfo {author} {\bibfnamefont {M.}~\bibnamefont
  {Oh}}, \bibinfo {author} {\bibfnamefont {K.~P.}\ \bibnamefont {Nuckolls}},
  \bibinfo {author} {\bibfnamefont {D.}~\bibnamefont {Wong}}, \bibinfo {author}
  {\bibfnamefont {R.~L.}\ \bibnamefont {Lee}}, \bibinfo {author} {\bibfnamefont
  {X.}~\bibnamefont {Liu}}, \bibinfo {author} {\bibfnamefont {K.}~\bibnamefont
  {Watanabe}}, \bibinfo {author} {\bibfnamefont {T.}~\bibnamefont {Taniguchi}},
  \ and\ \bibinfo {author} {\bibfnamefont {A.}~\bibnamefont {Yazdani}},\ }\href
  {\doibase 10.1038/s41586-021-04121-x} {\bibfield  {journal} {\bibinfo
  {journal} {Nature}\ }\textbf {\bibinfo {volume} {600}},\ \bibinfo {pages}
  {240} (\bibinfo {year} {2021})}\BibitemShut {NoStop}%
\bibitem [{\citenamefont {Park}\ \emph {et~al.}(2021)\citenamefont {Park},
  \citenamefont {Cao}, \citenamefont {Watanabe}, \citenamefont {Taniguchi},\
  and\ \citenamefont {Jarillo-Herrero}}]{Park2021}%
  \BibitemOpen
  \bibfield  {author} {\bibinfo {author} {\bibfnamefont {J.~M.}\ \bibnamefont
  {Park}}, \bibinfo {author} {\bibfnamefont {Y.}~\bibnamefont {Cao}}, \bibinfo
  {author} {\bibfnamefont {K.}~\bibnamefont {Watanabe}}, \bibinfo {author}
  {\bibfnamefont {T.}~\bibnamefont {Taniguchi}}, \ and\ \bibinfo {author}
  {\bibfnamefont {P.}~\bibnamefont {Jarillo-Herrero}},\ }\href {\doibase
  10.1038/s41586-021-03192-0} {\bibfield  {journal} {\bibinfo  {journal}
  {Nature}\ }\textbf {\bibinfo {volume} {590}},\ \bibinfo {pages} {249}
  (\bibinfo {year} {2021})}\BibitemShut {NoStop}%
\bibitem [{\citenamefont {Hao}\ \emph {et~al.}(2021)\citenamefont {Hao},
  \citenamefont {Zimmerman}, \citenamefont {Ledwith}, \citenamefont {Khalaf},
  \citenamefont {Najafabadi}, \citenamefont {Watanabe}, \citenamefont
  {Taniguchi}, \citenamefont {Vishwanath},\ and\ \citenamefont
  {Kim}}]{doi:10.1126/science.abg0399}%
  \BibitemOpen
  \bibfield  {author} {\bibinfo {author} {\bibfnamefont {Z.}~\bibnamefont
  {Hao}}, \bibinfo {author} {\bibfnamefont {A.~M.}\ \bibnamefont {Zimmerman}},
  \bibinfo {author} {\bibfnamefont {P.}~\bibnamefont {Ledwith}}, \bibinfo
  {author} {\bibfnamefont {E.}~\bibnamefont {Khalaf}}, \bibinfo {author}
  {\bibfnamefont {D.~H.}\ \bibnamefont {Najafabadi}}, \bibinfo {author}
  {\bibfnamefont {K.}~\bibnamefont {Watanabe}}, \bibinfo {author}
  {\bibfnamefont {T.}~\bibnamefont {Taniguchi}}, \bibinfo {author}
  {\bibfnamefont {A.}~\bibnamefont {Vishwanath}}, \ and\ \bibinfo {author}
  {\bibfnamefont {P.}~\bibnamefont {Kim}},\ }\href {\doibase
  10.1126/science.abg0399} {\bibfield  {journal} {\bibinfo  {journal}
  {Science}\ }\textbf {\bibinfo {volume} {371}},\ \bibinfo {pages} {1133}
  (\bibinfo {year} {2021})},\ \Eprint
  {http://arxiv.org/abs/https://www.science.org/doi/pdf/10.1126/science.abg0399}
  {https://www.science.org/doi/pdf/10.1126/science.abg0399} \BibitemShut
  {NoStop}%
\bibitem [{\citenamefont {Zhou}\ \emph {et~al.}(2022)\citenamefont {Zhou},
  \citenamefont {Holleis}, \citenamefont {Saito}, \citenamefont {Cohen},
  \citenamefont {Huynh}, \citenamefont {Patterson}, \citenamefont {Yang},
  \citenamefont {Taniguchi}, \citenamefont {Watanabe},\ and\ \citenamefont
  {Young}}]{doi:10.1126/science.abm8386}%
  \BibitemOpen
  \bibfield  {author} {\bibinfo {author} {\bibfnamefont {H.}~\bibnamefont
  {Zhou}}, \bibinfo {author} {\bibfnamefont {L.}~\bibnamefont {Holleis}},
  \bibinfo {author} {\bibfnamefont {Y.}~\bibnamefont {Saito}}, \bibinfo
  {author} {\bibfnamefont {L.}~\bibnamefont {Cohen}}, \bibinfo {author}
  {\bibfnamefont {W.}~\bibnamefont {Huynh}}, \bibinfo {author} {\bibfnamefont
  {C.~L.}\ \bibnamefont {Patterson}}, \bibinfo {author} {\bibfnamefont
  {F.}~\bibnamefont {Yang}}, \bibinfo {author} {\bibfnamefont {T.}~\bibnamefont
  {Taniguchi}}, \bibinfo {author} {\bibfnamefont {K.}~\bibnamefont {Watanabe}},
  \ and\ \bibinfo {author} {\bibfnamefont {A.~F.}\ \bibnamefont {Young}},\
  }\href {\doibase 10.1126/science.abm8386} {\bibfield  {journal} {\bibinfo
  {journal} {Science}\ }\textbf {\bibinfo {volume} {375}},\ \bibinfo {pages}
  {774} (\bibinfo {year} {2022})},\ \Eprint
  {http://arxiv.org/abs/https://www.science.org/doi/pdf/10.1126/science.abm8386}
  {https://www.science.org/doi/pdf/10.1126/science.abm8386} \BibitemShut
  {NoStop}%
\bibitem [{\citenamefont {Zhang}\ \emph {et~al.}(2023)\citenamefont {Zhang},
  \citenamefont {Polski}, \citenamefont {Thomson}, \citenamefont
  {Lantagne-Hurtubise}, \citenamefont {Lewandowski}, \citenamefont {Zhou},
  \citenamefont {Watanabe}, \citenamefont {Taniguchi}, \citenamefont {Alicea},\
  and\ \citenamefont {Nadj-Perge}}]{Zhang2023}%
  \BibitemOpen
  \bibfield  {author} {\bibinfo {author} {\bibfnamefont {Y.}~\bibnamefont
  {Zhang}}, \bibinfo {author} {\bibfnamefont {R.}~\bibnamefont {Polski}},
  \bibinfo {author} {\bibfnamefont {A.}~\bibnamefont {Thomson}}, \bibinfo
  {author} {\bibfnamefont {{\'E}.}~\bibnamefont {Lantagne-Hurtubise}}, \bibinfo
  {author} {\bibfnamefont {C.}~\bibnamefont {Lewandowski}}, \bibinfo {author}
  {\bibfnamefont {H.}~\bibnamefont {Zhou}}, \bibinfo {author} {\bibfnamefont
  {K.}~\bibnamefont {Watanabe}}, \bibinfo {author} {\bibfnamefont
  {T.}~\bibnamefont {Taniguchi}}, \bibinfo {author} {\bibfnamefont
  {J.}~\bibnamefont {Alicea}}, \ and\ \bibinfo {author} {\bibfnamefont
  {S.}~\bibnamefont {Nadj-Perge}},\ }\href {\doibase
  10.1038/s41586-022-05446-x} {\bibfield  {journal} {\bibinfo  {journal}
  {Nature}\ }\textbf {\bibinfo {volume} {613}},\ \bibinfo {pages} {268}
  (\bibinfo {year} {2023})}\BibitemShut {NoStop}%
\bibitem [{\citenamefont {Li}\ \emph {et~al.}(2024)\citenamefont {Li},
  \citenamefont {Xu}, \citenamefont {Li}, \citenamefont {Li}, \citenamefont
  {Li}, \citenamefont {Watanabe}, \citenamefont {Taniguchi}, \citenamefont
  {Tong}, \citenamefont {Shen}, \citenamefont {Lu}, \citenamefont {Jia},
  \citenamefont {Wu}, \citenamefont {Liu},\ and\ \citenamefont {Li}}]{Li2024}%
  \BibitemOpen
  \bibfield  {author} {\bibinfo {author} {\bibfnamefont {C.}~\bibnamefont
  {Li}}, \bibinfo {author} {\bibfnamefont {F.}~\bibnamefont {Xu}}, \bibinfo
  {author} {\bibfnamefont {B.}~\bibnamefont {Li}}, \bibinfo {author}
  {\bibfnamefont {J.}~\bibnamefont {Li}}, \bibinfo {author} {\bibfnamefont
  {G.}~\bibnamefont {Li}}, \bibinfo {author} {\bibfnamefont {K.}~\bibnamefont
  {Watanabe}}, \bibinfo {author} {\bibfnamefont {T.}~\bibnamefont {Taniguchi}},
  \bibinfo {author} {\bibfnamefont {B.}~\bibnamefont {Tong}}, \bibinfo {author}
  {\bibfnamefont {J.}~\bibnamefont {Shen}}, \bibinfo {author} {\bibfnamefont
  {L.}~\bibnamefont {Lu}}, \bibinfo {author} {\bibfnamefont {J.}~\bibnamefont
  {Jia}}, \bibinfo {author} {\bibfnamefont {F.}~\bibnamefont {Wu}}, \bibinfo
  {author} {\bibfnamefont {X.}~\bibnamefont {Liu}}, \ and\ \bibinfo {author}
  {\bibfnamefont {T.}~\bibnamefont {Li}},\ }\href {\doibase
  10.1038/s41586-024-07584-w} {\bibfield  {journal} {\bibinfo  {journal}
  {Nature}\ }\textbf {\bibinfo {volume} {631}},\ \bibinfo {pages} {300}
  (\bibinfo {year} {2024})}\BibitemShut {NoStop}%
\bibitem [{\citenamefont {Holleis}\ \emph {et~al.}(2023)\citenamefont
  {Holleis}, \citenamefont {Patterson}, \citenamefont {Zhang}, \citenamefont
  {Yoo}, \citenamefont {Zhou}, \citenamefont {Taniguchi}, \citenamefont
  {Watanabe}, \citenamefont {Nadj-Perge},\ and\ \citenamefont
  {Young}}]{holleis2023ising}%
  \BibitemOpen
  \bibfield  {author} {\bibinfo {author} {\bibfnamefont {L.}~\bibnamefont
  {Holleis}}, \bibinfo {author} {\bibfnamefont {C.~L.}\ \bibnamefont
  {Patterson}}, \bibinfo {author} {\bibfnamefont {Y.}~\bibnamefont {Zhang}},
  \bibinfo {author} {\bibfnamefont {H.~M.}\ \bibnamefont {Yoo}}, \bibinfo
  {author} {\bibfnamefont {H.}~\bibnamefont {Zhou}}, \bibinfo {author}
  {\bibfnamefont {T.}~\bibnamefont {Taniguchi}}, \bibinfo {author}
  {\bibfnamefont {K.}~\bibnamefont {Watanabe}}, \bibinfo {author}
  {\bibfnamefont {S.}~\bibnamefont {Nadj-Perge}}, \ and\ \bibinfo {author}
  {\bibfnamefont {A.~F.}\ \bibnamefont {Young}},\ }\href@noop {} {\bibfield
  {journal} {\bibinfo  {journal} {arXiv preprint arXiv:2303.00742}\ } (\bibinfo
  {year} {2023})}\BibitemShut {NoStop}%
\bibitem [{\citenamefont {Zhang}\ \emph {et~al.}(2024)\citenamefont {Zhang},
  \citenamefont {Shavit}, \citenamefont {Ma}, \citenamefont {Han},
  \citenamefont {Watanabe}, \citenamefont {Taniguchi}, \citenamefont {Hsieh},
  \citenamefont {Lewandowski}, \citenamefont {von Oppen}, \citenamefont {Oreg}
  \emph {et~al.}}]{zhang2024twist}%
  \BibitemOpen
  \bibfield  {author} {\bibinfo {author} {\bibfnamefont {Y.}~\bibnamefont
  {Zhang}}, \bibinfo {author} {\bibfnamefont {G.}~\bibnamefont {Shavit}},
  \bibinfo {author} {\bibfnamefont {H.}~\bibnamefont {Ma}}, \bibinfo {author}
  {\bibfnamefont {Y.}~\bibnamefont {Han}}, \bibinfo {author} {\bibfnamefont
  {K.}~\bibnamefont {Watanabe}}, \bibinfo {author} {\bibfnamefont
  {T.}~\bibnamefont {Taniguchi}}, \bibinfo {author} {\bibfnamefont
  {D.}~\bibnamefont {Hsieh}}, \bibinfo {author} {\bibfnamefont
  {C.}~\bibnamefont {Lewandowski}}, \bibinfo {author} {\bibfnamefont
  {F.}~\bibnamefont {von Oppen}}, \bibinfo {author} {\bibfnamefont
  {Y.}~\bibnamefont {Oreg}},  \emph {et~al.},\ }\href@noop {} {\bibfield
  {journal} {\bibinfo  {journal} {arXiv preprint arXiv:2408.10335}\ } (\bibinfo
  {year} {2024})}\BibitemShut {NoStop}%
\bibitem [{\citenamefont {Zhou}\ \emph {et~al.}(2021)\citenamefont {Zhou},
  \citenamefont {Xie}, \citenamefont {Ghazaryan}, \citenamefont {Holder},
  \citenamefont {Ehrets}, \citenamefont {Spanton}, \citenamefont {Taniguchi},
  \citenamefont {Watanabe}, \citenamefont {Berg}, \citenamefont {Serbyn},\ and\
  \citenamefont {Young}}]{Zhou2021}%
  \BibitemOpen
  \bibfield  {author} {\bibinfo {author} {\bibfnamefont {H.}~\bibnamefont
  {Zhou}}, \bibinfo {author} {\bibfnamefont {T.}~\bibnamefont {Xie}}, \bibinfo
  {author} {\bibfnamefont {A.}~\bibnamefont {Ghazaryan}}, \bibinfo {author}
  {\bibfnamefont {T.}~\bibnamefont {Holder}}, \bibinfo {author} {\bibfnamefont
  {J.~R.}\ \bibnamefont {Ehrets}}, \bibinfo {author} {\bibfnamefont {E.~M.}\
  \bibnamefont {Spanton}}, \bibinfo {author} {\bibfnamefont {T.}~\bibnamefont
  {Taniguchi}}, \bibinfo {author} {\bibfnamefont {K.}~\bibnamefont {Watanabe}},
  \bibinfo {author} {\bibfnamefont {E.}~\bibnamefont {Berg}}, \bibinfo {author}
  {\bibfnamefont {M.}~\bibnamefont {Serbyn}}, \ and\ \bibinfo {author}
  {\bibfnamefont {A.~F.}\ \bibnamefont {Young}},\ }\href {\doibase
  10.1038/s41586-021-03938-w} {\bibfield  {journal} {\bibinfo  {journal}
  {Nature}\ }\textbf {\bibinfo {volume} {598}},\ \bibinfo {pages} {429}
  (\bibinfo {year} {2021})}\BibitemShut {NoStop}%
\bibitem [{\citenamefont {Patterson}\ \emph {et~al.}(2024)\citenamefont
  {Patterson}, \citenamefont {Sheekey}, \citenamefont {Arp}, \citenamefont
  {Holleis}, \citenamefont {Koh}, \citenamefont {Choi}, \citenamefont {Xie},
  \citenamefont {Xu}, \citenamefont {Redekop}, \citenamefont {Babikyan} \emph
  {et~al.}}]{patterson2024superconductivity}%
  \BibitemOpen
  \bibfield  {author} {\bibinfo {author} {\bibfnamefont {C.~L.}\ \bibnamefont
  {Patterson}}, \bibinfo {author} {\bibfnamefont {O.~I.}\ \bibnamefont
  {Sheekey}}, \bibinfo {author} {\bibfnamefont {T.~B.}\ \bibnamefont {Arp}},
  \bibinfo {author} {\bibfnamefont {L.~F.}\ \bibnamefont {Holleis}}, \bibinfo
  {author} {\bibfnamefont {J.~M.}\ \bibnamefont {Koh}}, \bibinfo {author}
  {\bibfnamefont {Y.}~\bibnamefont {Choi}}, \bibinfo {author} {\bibfnamefont
  {T.}~\bibnamefont {Xie}}, \bibinfo {author} {\bibfnamefont {S.}~\bibnamefont
  {Xu}}, \bibinfo {author} {\bibfnamefont {E.}~\bibnamefont {Redekop}},
  \bibinfo {author} {\bibfnamefont {G.}~\bibnamefont {Babikyan}},  \emph
  {et~al.},\ }\href@noop {} {\bibfield  {journal} {\bibinfo  {journal} {arXiv
  preprint arXiv:2408.10190}\ } (\bibinfo {year} {2024})}\BibitemShut {NoStop}%
\bibitem [{\citenamefont {Yang}\ \emph {et~al.}(2024)\citenamefont {Yang},
  \citenamefont {Shi}, \citenamefont {Ye}, \citenamefont {Yoon}, \citenamefont
  {Lu}, \citenamefont {Kakani}, \citenamefont {Han}, \citenamefont {Seo},
  \citenamefont {Shi}, \citenamefont {Watanabe} \emph
  {et~al.}}]{yang2024diverse}%
  \BibitemOpen
  \bibfield  {author} {\bibinfo {author} {\bibfnamefont {J.}~\bibnamefont
  {Yang}}, \bibinfo {author} {\bibfnamefont {X.}~\bibnamefont {Shi}}, \bibinfo
  {author} {\bibfnamefont {S.}~\bibnamefont {Ye}}, \bibinfo {author}
  {\bibfnamefont {C.}~\bibnamefont {Yoon}}, \bibinfo {author} {\bibfnamefont
  {Z.}~\bibnamefont {Lu}}, \bibinfo {author} {\bibfnamefont {V.}~\bibnamefont
  {Kakani}}, \bibinfo {author} {\bibfnamefont {T.}~\bibnamefont {Han}},
  \bibinfo {author} {\bibfnamefont {J.}~\bibnamefont {Seo}}, \bibinfo {author}
  {\bibfnamefont {L.}~\bibnamefont {Shi}}, \bibinfo {author} {\bibfnamefont
  {K.}~\bibnamefont {Watanabe}},  \emph {et~al.},\ }\href@noop {} {\bibfield
  {journal} {\bibinfo  {journal} {arXiv preprint arXiv:2408.09906}\ } (\bibinfo
  {year} {2024})}\BibitemShut {NoStop}%
\bibitem [{\citenamefont {Wu}\ \emph {et~al.}(2018)\citenamefont {Wu},
  \citenamefont {MacDonald},\ and\ \citenamefont
  {Martin}}]{PhysRevLett.121.257001}%
  \BibitemOpen
  \bibfield  {author} {\bibinfo {author} {\bibfnamefont {F.}~\bibnamefont
  {Wu}}, \bibinfo {author} {\bibfnamefont {A.~H.}\ \bibnamefont {MacDonald}}, \
  and\ \bibinfo {author} {\bibfnamefont {I.}~\bibnamefont {Martin}},\ }\href
  {\doibase 10.1103/PhysRevLett.121.257001} {\bibfield  {journal} {\bibinfo
  {journal} {Phys. Rev. Lett.}\ }\textbf {\bibinfo {volume} {121}},\ \bibinfo
  {pages} {257001} (\bibinfo {year} {2018})}\BibitemShut {NoStop}%
\bibitem [{\citenamefont {Wu}\ \emph {et~al.}(2019)\citenamefont {Wu},
  \citenamefont {Hwang},\ and\ \citenamefont {Das~Sarma}}]{PhysRevB.99.165112}%
  \BibitemOpen
  \bibfield  {author} {\bibinfo {author} {\bibfnamefont {F.}~\bibnamefont
  {Wu}}, \bibinfo {author} {\bibfnamefont {E.}~\bibnamefont {Hwang}}, \ and\
  \bibinfo {author} {\bibfnamefont {S.}~\bibnamefont {Das~Sarma}},\ }\href
  {\doibase 10.1103/PhysRevB.99.165112} {\bibfield  {journal} {\bibinfo
  {journal} {Phys. Rev. B}\ }\textbf {\bibinfo {volume} {99}},\ \bibinfo
  {pages} {165112} (\bibinfo {year} {2019})}\BibitemShut {NoStop}%
\bibitem [{\citenamefont {Lian}\ \emph {et~al.}(2019)\citenamefont {Lian},
  \citenamefont {Wang},\ and\ \citenamefont
  {Bernevig}}]{PhysRevLett.122.257002}%
  \BibitemOpen
  \bibfield  {author} {\bibinfo {author} {\bibfnamefont {B.}~\bibnamefont
  {Lian}}, \bibinfo {author} {\bibfnamefont {Z.}~\bibnamefont {Wang}}, \ and\
  \bibinfo {author} {\bibfnamefont {B.~A.}\ \bibnamefont {Bernevig}},\ }\href
  {\doibase 10.1103/PhysRevLett.122.257002} {\bibfield  {journal} {\bibinfo
  {journal} {Phys. Rev. Lett.}\ }\textbf {\bibinfo {volume} {122}},\ \bibinfo
  {pages} {257002} (\bibinfo {year} {2019})}\BibitemShut {NoStop}%
\bibitem [{\citenamefont {Chou}\ \emph
  {et~al.}(2021{\natexlab{a}})\citenamefont {Chou}, \citenamefont {Wu},
  \citenamefont {Sau},\ and\ \citenamefont
  {Das~Sarma}}]{PhysRevLett.127.187001}%
  \BibitemOpen
  \bibfield  {author} {\bibinfo {author} {\bibfnamefont {Y.-Z.}\ \bibnamefont
  {Chou}}, \bibinfo {author} {\bibfnamefont {F.}~\bibnamefont {Wu}}, \bibinfo
  {author} {\bibfnamefont {J.~D.}\ \bibnamefont {Sau}}, \ and\ \bibinfo
  {author} {\bibfnamefont {S.}~\bibnamefont {Das~Sarma}},\ }\href {\doibase
  10.1103/PhysRevLett.127.187001} {\bibfield  {journal} {\bibinfo  {journal}
  {Phys. Rev. Lett.}\ }\textbf {\bibinfo {volume} {127}},\ \bibinfo {pages}
  {187001} (\bibinfo {year} {2021}{\natexlab{a}})}\BibitemShut {NoStop}%
\bibitem [{\citenamefont {Tuo}\ \emph {et~al.}(2024)\citenamefont {Tuo},
  \citenamefont {Li}, \citenamefont {Wu}, \citenamefont {Sun},\ and\
  \citenamefont {Yao}}]{tuo2024theory}%
  \BibitemOpen
  \bibfield  {author} {\bibinfo {author} {\bibfnamefont {C.}~\bibnamefont
  {Tuo}}, \bibinfo {author} {\bibfnamefont {M.-R.}\ \bibnamefont {Li}},
  \bibinfo {author} {\bibfnamefont {Z.}~\bibnamefont {Wu}}, \bibinfo {author}
  {\bibfnamefont {W.}~\bibnamefont {Sun}}, \ and\ \bibinfo {author}
  {\bibfnamefont {H.}~\bibnamefont {Yao}},\ }\href@noop {} {\bibfield
  {journal} {\bibinfo  {journal} {arXiv:2409.06779}\ } (\bibinfo {year}
  {2024})}\BibitemShut {NoStop}%
\bibitem [{\citenamefont {Chou}\ \emph
  {et~al.}(2022{\natexlab{a}})\citenamefont {Chou}, \citenamefont {Wu},
  \citenamefont {Sau},\ and\ \citenamefont {Das~Sarma}}]{PhysRevB.106.024507}%
  \BibitemOpen
  \bibfield  {author} {\bibinfo {author} {\bibfnamefont {Y.-Z.}\ \bibnamefont
  {Chou}}, \bibinfo {author} {\bibfnamefont {F.}~\bibnamefont {Wu}}, \bibinfo
  {author} {\bibfnamefont {J.~D.}\ \bibnamefont {Sau}}, \ and\ \bibinfo
  {author} {\bibfnamefont {S.}~\bibnamefont {Das~Sarma}},\ }\href {\doibase
  10.1103/PhysRevB.106.024507} {\bibfield  {journal} {\bibinfo  {journal}
  {Phys. Rev. B}\ }\textbf {\bibinfo {volume} {106}},\ \bibinfo {pages}
  {024507} (\bibinfo {year} {2022}{\natexlab{a}})}\BibitemShut {NoStop}%
\bibitem [{\citenamefont {Chou}\ \emph
  {et~al.}(2022{\natexlab{b}})\citenamefont {Chou}, \citenamefont {Wu},
  \citenamefont {Sau},\ and\ \citenamefont {Das~Sarma}}]{PhysRevB.105.L100503}%
  \BibitemOpen
  \bibfield  {author} {\bibinfo {author} {\bibfnamefont {Y.-Z.}\ \bibnamefont
  {Chou}}, \bibinfo {author} {\bibfnamefont {F.}~\bibnamefont {Wu}}, \bibinfo
  {author} {\bibfnamefont {J.~D.}\ \bibnamefont {Sau}}, \ and\ \bibinfo
  {author} {\bibfnamefont {S.}~\bibnamefont {Das~Sarma}},\ }\href {\doibase
  10.1103/PhysRevB.105.L100503} {\bibfield  {journal} {\bibinfo  {journal}
  {Phys. Rev. B}\ }\textbf {\bibinfo {volume} {105}},\ \bibinfo {pages}
  {L100503} (\bibinfo {year} {2022}{\natexlab{b}})}\BibitemShut {NoStop}%
\bibitem [{\citenamefont {Chou}\ \emph
  {et~al.}(2024{\natexlab{a}})\citenamefont {Chou}, \citenamefont {Tan},
  \citenamefont {Wu},\ and\ \citenamefont {Das~Sarma}}]{PhysRevB.110.L041108}%
  \BibitemOpen
  \bibfield  {author} {\bibinfo {author} {\bibfnamefont {Y.-Z.}\ \bibnamefont
  {Chou}}, \bibinfo {author} {\bibfnamefont {Y.}~\bibnamefont {Tan}}, \bibinfo
  {author} {\bibfnamefont {F.}~\bibnamefont {Wu}}, \ and\ \bibinfo {author}
  {\bibfnamefont {S.}~\bibnamefont {Das~Sarma}},\ }\href {\doibase
  10.1103/PhysRevB.110.L041108} {\bibfield  {journal} {\bibinfo  {journal}
  {Phys. Rev. B}\ }\textbf {\bibinfo {volume} {110}},\ \bibinfo {pages}
  {L041108} (\bibinfo {year} {2024}{\natexlab{a}})}\BibitemShut {NoStop}%
\bibitem [{\citenamefont {Isobe}\ \emph {et~al.}(2018)\citenamefont {Isobe},
  \citenamefont {Yuan},\ and\ \citenamefont {Fu}}]{PhysRevX.8.041041}%
  \BibitemOpen
  \bibfield  {author} {\bibinfo {author} {\bibfnamefont {H.}~\bibnamefont
  {Isobe}}, \bibinfo {author} {\bibfnamefont {N.~F.~Q.}\ \bibnamefont {Yuan}},
  \ and\ \bibinfo {author} {\bibfnamefont {L.}~\bibnamefont {Fu}},\ }\href
  {\doibase 10.1103/PhysRevX.8.041041} {\bibfield  {journal} {\bibinfo
  {journal} {Phys. Rev. X}\ }\textbf {\bibinfo {volume} {8}},\ \bibinfo {pages}
  {041041} (\bibinfo {year} {2018})}\BibitemShut {NoStop}%
\bibitem [{\citenamefont {Yuan}\ \emph {et~al.}(2019)\citenamefont {Yuan},
  \citenamefont {Isobe},\ and\ \citenamefont {Fu}}]{Yuan2019}%
  \BibitemOpen
  \bibfield  {author} {\bibinfo {author} {\bibfnamefont {N.~F.~Q.}\
  \bibnamefont {Yuan}}, \bibinfo {author} {\bibfnamefont {H.}~\bibnamefont
  {Isobe}}, \ and\ \bibinfo {author} {\bibfnamefont {L.}~\bibnamefont {Fu}},\
  }\href {\doibase 10.1038/s41467-019-13670-9} {\bibfield  {journal} {\bibinfo
  {journal} {Nature Communications}\ }\textbf {\bibinfo {volume} {10}},\
  \bibinfo {pages} {5769} (\bibinfo {year} {2019})}\BibitemShut {NoStop}%
\bibitem [{\citenamefont {Lu}\ \emph {et~al.}(2022)\citenamefont {Lu},
  \citenamefont {Wang}, \citenamefont {Chatterjee},\ and\ \citenamefont
  {You}}]{PhysRevB.106.155115}%
  \BibitemOpen
  \bibfield  {author} {\bibinfo {author} {\bibfnamefont {D.-C.}\ \bibnamefont
  {Lu}}, \bibinfo {author} {\bibfnamefont {T.}~\bibnamefont {Wang}}, \bibinfo
  {author} {\bibfnamefont {S.}~\bibnamefont {Chatterjee}}, \ and\ \bibinfo
  {author} {\bibfnamefont {Y.-Z.}\ \bibnamefont {You}},\ }\href {\doibase
  10.1103/PhysRevB.106.155115} {\bibfield  {journal} {\bibinfo  {journal}
  {Phys. Rev. B}\ }\textbf {\bibinfo {volume} {106}},\ \bibinfo {pages}
  {155115} (\bibinfo {year} {2022})}\BibitemShut {NoStop}%
\bibitem [{\citenamefont {Ghazaryan}\ \emph {et~al.}(2021)\citenamefont
  {Ghazaryan}, \citenamefont {Holder}, \citenamefont {Serbyn},\ and\
  \citenamefont {Berg}}]{PhysRevLett.127.247001}%
  \BibitemOpen
  \bibfield  {author} {\bibinfo {author} {\bibfnamefont {A.}~\bibnamefont
  {Ghazaryan}}, \bibinfo {author} {\bibfnamefont {T.}~\bibnamefont {Holder}},
  \bibinfo {author} {\bibfnamefont {M.}~\bibnamefont {Serbyn}}, \ and\ \bibinfo
  {author} {\bibfnamefont {E.}~\bibnamefont {Berg}},\ }\href {\doibase
  10.1103/PhysRevLett.127.247001} {\bibfield  {journal} {\bibinfo  {journal}
  {Phys. Rev. Lett.}\ }\textbf {\bibinfo {volume} {127}},\ \bibinfo {pages}
  {247001} (\bibinfo {year} {2021})}\BibitemShut {NoStop}%
\bibitem [{\citenamefont {Cea}\ \emph {et~al.}(2022)\citenamefont {Cea},
  \citenamefont {Pantale\'on}, \citenamefont {Phong},\ and\ \citenamefont
  {Guinea}}]{PhysRevB.105.075432}%
  \BibitemOpen
  \bibfield  {author} {\bibinfo {author} {\bibfnamefont {T.}~\bibnamefont
  {Cea}}, \bibinfo {author} {\bibfnamefont {P.~A.}\ \bibnamefont
  {Pantale\'on}}, \bibinfo {author} {\bibfnamefont {V.~o.~T.}\ \bibnamefont
  {Phong}}, \ and\ \bibinfo {author} {\bibfnamefont {F.}~\bibnamefont
  {Guinea}},\ }\href {\doibase 10.1103/PhysRevB.105.075432} {\bibfield
  {journal} {\bibinfo  {journal} {Phys. Rev. B}\ }\textbf {\bibinfo {volume}
  {105}},\ \bibinfo {pages} {075432} (\bibinfo {year} {2022})}\BibitemShut
  {NoStop}%
\bibitem [{\citenamefont {Li}\ \emph {et~al.}(2023)\citenamefont {Li},
  \citenamefont {Kuang}, \citenamefont {Jimeno-Pozo}, \citenamefont
  {Sainz-Cruz}, \citenamefont {Zhan}, \citenamefont {Yuan},\ and\ \citenamefont
  {Guinea}}]{PhysRevB.108.045404}%
  \BibitemOpen
  \bibfield  {author} {\bibinfo {author} {\bibfnamefont {Z.}~\bibnamefont
  {Li}}, \bibinfo {author} {\bibfnamefont {X.}~\bibnamefont {Kuang}}, \bibinfo
  {author} {\bibfnamefont {A.}~\bibnamefont {Jimeno-Pozo}}, \bibinfo {author}
  {\bibfnamefont {H.}~\bibnamefont {Sainz-Cruz}}, \bibinfo {author}
  {\bibfnamefont {Z.}~\bibnamefont {Zhan}}, \bibinfo {author} {\bibfnamefont
  {S.}~\bibnamefont {Yuan}}, \ and\ \bibinfo {author} {\bibfnamefont
  {F.}~\bibnamefont {Guinea}},\ }\href {\doibase 10.1103/PhysRevB.108.045404}
  {\bibfield  {journal} {\bibinfo  {journal} {Phys. Rev. B}\ }\textbf {\bibinfo
  {volume} {108}},\ \bibinfo {pages} {045404} (\bibinfo {year}
  {2023})}\BibitemShut {NoStop}%
\bibitem [{\citenamefont {Jimeno-Pozo}\ \emph {et~al.}(2023)\citenamefont
  {Jimeno-Pozo}, \citenamefont {Sainz-Cruz}, \citenamefont {Cea}, \citenamefont
  {Pantale\'on},\ and\ \citenamefont {Guinea}}]{PhysRevB.107.L161106}%
  \BibitemOpen
  \bibfield  {author} {\bibinfo {author} {\bibfnamefont {A.}~\bibnamefont
  {Jimeno-Pozo}}, \bibinfo {author} {\bibfnamefont {H.}~\bibnamefont
  {Sainz-Cruz}}, \bibinfo {author} {\bibfnamefont {T.}~\bibnamefont {Cea}},
  \bibinfo {author} {\bibfnamefont {P.~A.}\ \bibnamefont {Pantale\'on}}, \ and\
  \bibinfo {author} {\bibfnamefont {F.}~\bibnamefont {Guinea}},\ }\href
  {\doibase 10.1103/PhysRevB.107.L161106} {\bibfield  {journal} {\bibinfo
  {journal} {Phys. Rev. B}\ }\textbf {\bibinfo {volume} {107}},\ \bibinfo
  {pages} {L161106} (\bibinfo {year} {2023})}\BibitemShut {NoStop}%
\bibitem [{\citenamefont {Guerci}\ \emph {et~al.}(2024)\citenamefont {Guerci},
  \citenamefont {Kaplan}, \citenamefont {Ingham}, \citenamefont {Pixley},\ and\
  \citenamefont {Millis}}]{guerci2024topological}%
  \BibitemOpen
  \bibfield  {author} {\bibinfo {author} {\bibfnamefont {D.}~\bibnamefont
  {Guerci}}, \bibinfo {author} {\bibfnamefont {D.}~\bibnamefont {Kaplan}},
  \bibinfo {author} {\bibfnamefont {J.}~\bibnamefont {Ingham}}, \bibinfo
  {author} {\bibfnamefont {J.}~\bibnamefont {Pixley}}, \ and\ \bibinfo {author}
  {\bibfnamefont {A.~J.}\ \bibnamefont {Millis}},\ }\href@noop {} {\bibfield
  {journal} {\bibinfo  {journal} {arXiv:2408.16075}\ } (\bibinfo {year}
  {2024})}\BibitemShut {NoStop}%
\bibitem [{\citenamefont {Qin}\ \emph {et~al.}(2024)\citenamefont {Qin},
  \citenamefont {Qiu},\ and\ \citenamefont {Wu}}]{qin2024kohn}%
  \BibitemOpen
  \bibfield  {author} {\bibinfo {author} {\bibfnamefont {W.}~\bibnamefont
  {Qin}}, \bibinfo {author} {\bibfnamefont {W.-X.}\ \bibnamefont {Qiu}}, \ and\
  \bibinfo {author} {\bibfnamefont {F.}~\bibnamefont {Wu}},\ }\href@noop {}
  {\bibfield  {journal} {\bibinfo  {journal} {arXiv:2409.16114}\ } (\bibinfo
  {year} {2024})}\BibitemShut {NoStop}%
\bibitem [{\citenamefont {Xu}\ and\ \citenamefont
  {Balents}(2018)}]{PhysRevLett.121.087001}%
  \BibitemOpen
  \bibfield  {author} {\bibinfo {author} {\bibfnamefont {C.}~\bibnamefont
  {Xu}}\ and\ \bibinfo {author} {\bibfnamefont {L.}~\bibnamefont {Balents}},\
  }\href {\doibase 10.1103/PhysRevLett.121.087001} {\bibfield  {journal}
  {\bibinfo  {journal} {Phys. Rev. Lett.}\ }\textbf {\bibinfo {volume} {121}},\
  \bibinfo {pages} {087001} (\bibinfo {year} {2018})}\BibitemShut {NoStop}%
\bibitem [{\citenamefont {You}\ and\ \citenamefont
  {Vishwanath}(2019)}]{You2019}%
  \BibitemOpen
  \bibfield  {author} {\bibinfo {author} {\bibfnamefont {Y.-Z.}\ \bibnamefont
  {You}}\ and\ \bibinfo {author} {\bibfnamefont {A.}~\bibnamefont
  {Vishwanath}},\ }\href {\doibase 10.1038/s41535-019-0153-4} {\bibfield
  {journal} {\bibinfo  {journal} {npj Quantum Materials}\ }\textbf {\bibinfo
  {volume} {4}},\ \bibinfo {pages} {16} (\bibinfo {year} {2019})}\BibitemShut
  {NoStop}%
\bibitem [{\citenamefont {Lee}\ \emph {et~al.}(2019)\citenamefont {Lee},
  \citenamefont {Khalaf}, \citenamefont {Liu}, \citenamefont {Liu},
  \citenamefont {Hao}, \citenamefont {Kim},\ and\ \citenamefont
  {Vishwanath}}]{Lee2019}%
  \BibitemOpen
  \bibfield  {author} {\bibinfo {author} {\bibfnamefont {J.~Y.}\ \bibnamefont
  {Lee}}, \bibinfo {author} {\bibfnamefont {E.}~\bibnamefont {Khalaf}},
  \bibinfo {author} {\bibfnamefont {S.}~\bibnamefont {Liu}}, \bibinfo {author}
  {\bibfnamefont {X.}~\bibnamefont {Liu}}, \bibinfo {author} {\bibfnamefont
  {Z.}~\bibnamefont {Hao}}, \bibinfo {author} {\bibfnamefont {P.}~\bibnamefont
  {Kim}}, \ and\ \bibinfo {author} {\bibfnamefont {A.}~\bibnamefont
  {Vishwanath}},\ }\href {\doibase 10.1038/s41467-019-12981-1} {\bibfield
  {journal} {\bibinfo  {journal} {Nature Communications}\ }\textbf {\bibinfo
  {volume} {10}},\ \bibinfo {pages} {5333} (\bibinfo {year}
  {2019})}\BibitemShut {NoStop}%
\bibitem [{\citenamefont {Chatterjee}\ \emph {et~al.}(2022)\citenamefont
  {Chatterjee}, \citenamefont {Wang}, \citenamefont {Berg},\ and\ \citenamefont
  {Zaletel}}]{Chatterjee2022}%
  \BibitemOpen
  \bibfield  {author} {\bibinfo {author} {\bibfnamefont {S.}~\bibnamefont
  {Chatterjee}}, \bibinfo {author} {\bibfnamefont {T.}~\bibnamefont {Wang}},
  \bibinfo {author} {\bibfnamefont {E.}~\bibnamefont {Berg}}, \ and\ \bibinfo
  {author} {\bibfnamefont {M.~P.}\ \bibnamefont {Zaletel}},\ }\href {\doibase
  10.1038/s41467-022-33561-w} {\bibfield  {journal} {\bibinfo  {journal}
  {Nature Communications}\ }\textbf {\bibinfo {volume} {13}},\ \bibinfo {pages}
  {6013} (\bibinfo {year} {2022})}\BibitemShut {NoStop}%
\bibitem [{\citenamefont {Szab\'o}\ and\ \citenamefont
  {Roy}(2022)}]{PhysRevB.105.L081407}%
  \BibitemOpen
  \bibfield  {author} {\bibinfo {author} {\bibfnamefont {A.~L.}\ \bibnamefont
  {Szab\'o}}\ and\ \bibinfo {author} {\bibfnamefont {B.}~\bibnamefont {Roy}},\
  }\href {\doibase 10.1103/PhysRevB.105.L081407} {\bibfield  {journal}
  {\bibinfo  {journal} {Phys. Rev. B}\ }\textbf {\bibinfo {volume} {105}},\
  \bibinfo {pages} {L081407} (\bibinfo {year} {2022})}\BibitemShut {NoStop}%
\bibitem [{\citenamefont {Christos}\ \emph {et~al.}(2022)\citenamefont
  {Christos}, \citenamefont {Sachdev},\ and\ \citenamefont
  {Scheurer}}]{PhysRevX.12.021018}%
  \BibitemOpen
  \bibfield  {author} {\bibinfo {author} {\bibfnamefont {M.}~\bibnamefont
  {Christos}}, \bibinfo {author} {\bibfnamefont {S.}~\bibnamefont {Sachdev}}, \
  and\ \bibinfo {author} {\bibfnamefont {M.~S.}\ \bibnamefont {Scheurer}},\
  }\href {\doibase 10.1103/PhysRevX.12.021018} {\bibfield  {journal} {\bibinfo
  {journal} {Phys. Rev. X}\ }\textbf {\bibinfo {volume} {12}},\ \bibinfo
  {pages} {021018} (\bibinfo {year} {2022})}\BibitemShut {NoStop}%
\bibitem [{\citenamefont {Chou}\ \emph
  {et~al.}(2021{\natexlab{b}})\citenamefont {Chou}, \citenamefont {Wu},
  \citenamefont {Sau},\ and\ \citenamefont
  {Das~Sarma}}]{PhysRevLett.127.217001}%
  \BibitemOpen
  \bibfield  {author} {\bibinfo {author} {\bibfnamefont {Y.-Z.}\ \bibnamefont
  {Chou}}, \bibinfo {author} {\bibfnamefont {F.}~\bibnamefont {Wu}}, \bibinfo
  {author} {\bibfnamefont {J.~D.}\ \bibnamefont {Sau}}, \ and\ \bibinfo
  {author} {\bibfnamefont {S.}~\bibnamefont {Das~Sarma}},\ }\href {\doibase
  10.1103/PhysRevLett.127.217001} {\bibfield  {journal} {\bibinfo  {journal}
  {Phys. Rev. Lett.}\ }\textbf {\bibinfo {volume} {127}},\ \bibinfo {pages}
  {217001} (\bibinfo {year} {2021}{\natexlab{b}})}\BibitemShut {NoStop}%
\bibitem [{\citenamefont {Dong}\ \emph {et~al.}(2023)\citenamefont {Dong},
  \citenamefont {Levitov},\ and\ \citenamefont
  {Chubukov}}]{dong2023superconductivity}%
  \BibitemOpen
  \bibfield  {author} {\bibinfo {author} {\bibfnamefont {Z.}~\bibnamefont
  {Dong}}, \bibinfo {author} {\bibfnamefont {L.}~\bibnamefont {Levitov}}, \
  and\ \bibinfo {author} {\bibfnamefont {A.~V.}\ \bibnamefont {Chubukov}},\
  }\href@noop {} {\bibfield  {journal} {\bibinfo  {journal} {arXiv:2306.11005}\
  } (\bibinfo {year} {2023})}\BibitemShut {NoStop}%
\bibitem [{\citenamefont {Dong}\ \emph {et~al.}(2024)\citenamefont {Dong},
  \citenamefont {Lantagne-Hurtubise},\ and\ \citenamefont
  {Alicea}}]{dong2024superconductivity}%
  \BibitemOpen
  \bibfield  {author} {\bibinfo {author} {\bibfnamefont {Z.}~\bibnamefont
  {Dong}}, \bibinfo {author} {\bibfnamefont {{\'E}.}~\bibnamefont
  {Lantagne-Hurtubise}}, \ and\ \bibinfo {author} {\bibfnamefont
  {J.}~\bibnamefont {Alicea}},\ }\href@noop {} {\bibfield  {journal} {\bibinfo
  {journal} {arXiv:2406.17036}\ } (\bibinfo {year} {2024})}\BibitemShut
  {NoStop}%
\bibitem [{\citenamefont {Zhou}\ and\ \citenamefont
  {Zhang}(2023)}]{PhysRevB.108.155111}%
  \BibitemOpen
  \bibfield  {author} {\bibinfo {author} {\bibfnamefont {B.}~\bibnamefont
  {Zhou}}\ and\ \bibinfo {author} {\bibfnamefont {Y.-H.}\ \bibnamefont
  {Zhang}},\ }\href {\doibase 10.1103/PhysRevB.108.155111} {\bibfield
  {journal} {\bibinfo  {journal} {Phys. Rev. B}\ }\textbf {\bibinfo {volume}
  {108}},\ \bibinfo {pages} {155111} (\bibinfo {year} {2023})}\BibitemShut
  {NoStop}%
\bibitem [{\citenamefont {Kim}\ \emph {et~al.}(2024{\natexlab{a}})\citenamefont
  {Kim}, \citenamefont {Mendez-Valderrama}, \citenamefont {Wang},\ and\
  \citenamefont {Chowdhury}}]{kim2024theory}%
  \BibitemOpen
  \bibfield  {author} {\bibinfo {author} {\bibfnamefont {S.}~\bibnamefont
  {Kim}}, \bibinfo {author} {\bibfnamefont {J.~F.}\ \bibnamefont
  {Mendez-Valderrama}}, \bibinfo {author} {\bibfnamefont {X.}~\bibnamefont
  {Wang}}, \ and\ \bibinfo {author} {\bibfnamefont {D.}~\bibnamefont
  {Chowdhury}},\ }\href@noop {} {\bibfield  {journal} {\bibinfo  {journal}
  {arXiv:2406.03525}\ } (\bibinfo {year} {2024}{\natexlab{a}})}\BibitemShut
  {NoStop}%
\bibitem [{\citenamefont {Khalaf}\ \emph {et~al.}(2021)\citenamefont {Khalaf},
  \citenamefont {Chatterjee}, \citenamefont {Bultinck}, \citenamefont
  {Zaletel},\ and\ \citenamefont {Vishwanath}}]{doi:10.1126/sciadv.abf5299}%
  \BibitemOpen
  \bibfield  {author} {\bibinfo {author} {\bibfnamefont {E.}~\bibnamefont
  {Khalaf}}, \bibinfo {author} {\bibfnamefont {S.}~\bibnamefont {Chatterjee}},
  \bibinfo {author} {\bibfnamefont {N.}~\bibnamefont {Bultinck}}, \bibinfo
  {author} {\bibfnamefont {M.~P.}\ \bibnamefont {Zaletel}}, \ and\ \bibinfo
  {author} {\bibfnamefont {A.}~\bibnamefont {Vishwanath}},\ }\href {\doibase
  10.1126/sciadv.abf5299} {\bibfield  {journal} {\bibinfo  {journal} {Science
  Advances}\ }\textbf {\bibinfo {volume} {7}},\ \bibinfo {pages} {eabf5299}
  (\bibinfo {year} {2021})},\ \Eprint
  {http://arxiv.org/abs/https://www.science.org/doi/pdf/10.1126/sciadv.abf5299}
  {https://www.science.org/doi/pdf/10.1126/sciadv.abf5299} \BibitemShut
  {NoStop}%
\bibitem [{\citenamefont {Shi}\ and\ \citenamefont
  {Senthil}(2024)}]{shi2024doping}%
  \BibitemOpen
  \bibfield  {author} {\bibinfo {author} {\bibfnamefont {Z.~D.}\ \bibnamefont
  {Shi}}\ and\ \bibinfo {author} {\bibfnamefont {T.}~\bibnamefont {Senthil}},\
  }\href@noop {} {\bibfield  {journal} {\bibinfo  {journal} {arXiv:2409.20567}\
  } (\bibinfo {year} {2024})}\BibitemShut {NoStop}%
\bibitem [{\citenamefont {Kim}\ \emph {et~al.}(2024{\natexlab{b}})\citenamefont
  {Kim}, \citenamefont {Timmel}, \citenamefont {Ju},\ and\ \citenamefont
  {Wen}}]{Kim2024Topological}%
  \BibitemOpen
  \bibfield  {author} {\bibinfo {author} {\bibfnamefont {M.}~\bibnamefont
  {Kim}}, \bibinfo {author} {\bibfnamefont {A.}~\bibnamefont {Timmel}},
  \bibinfo {author} {\bibfnamefont {L.}~\bibnamefont {Ju}}, \ and\ \bibinfo
  {author} {\bibfnamefont {X.-G.}\ \bibnamefont {Wen}},\ }\href@noop {}
  {\bibfield  {journal} {\bibinfo  {journal} {arXiv preprint arXiv:2409.18067}\
  } (\bibinfo {year} {2024}{\natexlab{b}})}\BibitemShut {NoStop}%
\bibitem [{\citenamefont {Chou}\ \emph
  {et~al.}(2024{\natexlab{b}})\citenamefont {Chou}, \citenamefont {Zhu},\ and\
  \citenamefont {Sarma}}]{Chou2024intravalley}%
  \BibitemOpen
  \bibfield  {author} {\bibinfo {author} {\bibfnamefont {Y.-Z.}\ \bibnamefont
  {Chou}}, \bibinfo {author} {\bibfnamefont {J.}~\bibnamefont {Zhu}}, \ and\
  \bibinfo {author} {\bibfnamefont {S.~D.}\ \bibnamefont {Sarma}},\ }\href@noop
  {} {\bibfield  {journal} {\bibinfo  {journal} {arXiv preprint
  arXiv:2409.06701}\ } (\bibinfo {year} {2024}{\natexlab{b}})}\BibitemShut
  {NoStop}%
\bibitem [{\citenamefont {Geier}\ \emph {et~al.}(2024)\citenamefont {Geier},
  \citenamefont {Davydova},\ and\ \citenamefont {Fu}}]{Geier2024Chiral}%
  \BibitemOpen
  \bibfield  {author} {\bibinfo {author} {\bibfnamefont {M.}~\bibnamefont
  {Geier}}, \bibinfo {author} {\bibfnamefont {M.}~\bibnamefont {Davydova}}, \
  and\ \bibinfo {author} {\bibfnamefont {L.}~\bibnamefont {Fu}},\ }\href
  {https://arxiv.org/abs/2409.13829} {\enquote {\bibinfo {title} {Chiral and
  topological superconductivity in isospin polarized multilayer graphene},}\ }
  (\bibinfo {year} {2024}),\ \Eprint {http://arxiv.org/abs/2409.13829}
  {arXiv:2409.13829 [cond-mat.supr-con]} \BibitemShut {NoStop}%
\bibitem [{Note1()}]{Note1}%
  \BibitemOpen
  \bibinfo {note} {The real Berezinskii-Kosterlitz-Thouless (BKT) critical
  temperature is different from mean field $T_c$ and is decided by phase
  stiffness.}\BibitemShut {Stop}%
\bibitem [{\citenamefont {Fulde}\ and\ \citenamefont
  {Ferrell}(1964)}]{PhysRev.135.A550}%
  \BibitemOpen
  \bibfield  {author} {\bibinfo {author} {\bibfnamefont {P.}~\bibnamefont
  {Fulde}}\ and\ \bibinfo {author} {\bibfnamefont {R.~A.}\ \bibnamefont
  {Ferrell}},\ }\href {\doibase 10.1103/PhysRev.135.A550} {\bibfield  {journal}
  {\bibinfo  {journal} {Phys. Rev.}\ }\textbf {\bibinfo {volume} {135}},\
  \bibinfo {pages} {A550} (\bibinfo {year} {1964})}\BibitemShut {NoStop}%
\bibitem [{\citenamefont {Larkin}\ and\ \citenamefont
  {Ovchinnikov}(1964)}]{Larkin:1964wok}%
  \BibitemOpen
  \bibfield  {author} {\bibinfo {author} {\bibfnamefont {A.~I.}\ \bibnamefont
  {Larkin}}\ and\ \bibinfo {author} {\bibfnamefont {Y.~N.}\ \bibnamefont
  {Ovchinnikov}},\ }\href@noop {} {\bibfield  {journal} {\bibinfo  {journal}
  {Zh. Eksp. Teor. Fiz.}\ }\textbf {\bibinfo {volume} {47}},\ \bibinfo {pages}
  {1136} (\bibinfo {year} {1964})}\BibitemShut {NoStop}%
\bibitem [{\citenamefont {Casalbuoni}\ and\ \citenamefont
  {Nardulli}(2004)}]{RevModPhys.76.263}%
  \BibitemOpen
  \bibfield  {author} {\bibinfo {author} {\bibfnamefont {R.}~\bibnamefont
  {Casalbuoni}}\ and\ \bibinfo {author} {\bibfnamefont {G.}~\bibnamefont
  {Nardulli}},\ }\href {\doibase 10.1103/RevModPhys.76.263} {\bibfield
  {journal} {\bibinfo  {journal} {Rev. Mod. Phys.}\ }\textbf {\bibinfo {volume}
  {76}},\ \bibinfo {pages} {263} (\bibinfo {year} {2004})}\BibitemShut
  {NoStop}%
\bibitem [{\citenamefont {Min}\ and\ \citenamefont
  {MacDonald}(2008)}]{min2008electronic}%
  \BibitemOpen
  \bibfield  {author} {\bibinfo {author} {\bibfnamefont {H.}~\bibnamefont
  {Min}}\ and\ \bibinfo {author} {\bibfnamefont {A.~H.}\ \bibnamefont
  {MacDonald}},\ }\href@noop {} {\bibfield  {journal} {\bibinfo  {journal}
  {Progress of Theoretical Physics Supplement}\ }\textbf {\bibinfo {volume}
  {176}},\ \bibinfo {pages} {227} (\bibinfo {year} {2008})}\BibitemShut
  {NoStop}%
\bibitem [{\citenamefont {Ghazaryan}\ \emph {et~al.}(2023)\citenamefont
  {Ghazaryan}, \citenamefont {Holder}, \citenamefont {Berg},\ and\
  \citenamefont {Serbyn}}]{PhysRevB.107.104502}%
  \BibitemOpen
  \bibfield  {author} {\bibinfo {author} {\bibfnamefont {A.}~\bibnamefont
  {Ghazaryan}}, \bibinfo {author} {\bibfnamefont {T.}~\bibnamefont {Holder}},
  \bibinfo {author} {\bibfnamefont {E.}~\bibnamefont {Berg}}, \ and\ \bibinfo
  {author} {\bibfnamefont {M.}~\bibnamefont {Serbyn}},\ }\href {\doibase
  10.1103/PhysRevB.107.104502} {\bibfield  {journal} {\bibinfo  {journal}
  {Phys. Rev. B}\ }\textbf {\bibinfo {volume} {107}},\ \bibinfo {pages}
  {104502} (\bibinfo {year} {2023})}\BibitemShut {NoStop}%
\bibitem [{Note2()}]{Note2}%
  \BibitemOpen
  \bibinfo {note} {The maximal $k_F$ is defined as the largest $|\protect
  \mathbf k|$ on the Fermi surface.}\BibitemShut {Stop}%
\end{thebibliography}%
\onecolumngrid
\appendix
\section{Details about the mean-field theory}
Considering the case finite momentum pairing with momentum ${\bf Q}$, we define the mean-field ansatz
\begin{align}
    \Delta_Q(k)=\frac{1}{A}\sum_{k_2}V(k_2-k_1)\Lambda_{k_1+\frac{Q}{2},k_2-k_1}\Lambda_{-k_1+\frac{Q}{2},k_1-k_2}\langle c_{-k_2+\frac{Q}{2}}c_{k_2+\frac{Q}{2}}\rangle,
\end{align}
then the mean-field Hamiltonian becomes,
\begin{align}
    H_{MF}=&\sum_{k}\epsilon_{k} c^\dag_k c_k+\Delta_Q(k)c^\dag_{k_1+\frac{Q}{2}}c^\dag_{-k_1+\frac{Q}{2}}+h.c.\nonumber\\
    =&\sum_{k>0}\epsilon_{k+\frac{Q}{2}} c^\dag_k c_k+\epsilon_{{-k+\frac{Q}{2}}} c^\dag_{-k+\frac{Q}{2}} c_{-k+\frac{Q}{2}}+\frac{1}{2}\Delta_Q(k)c^\dag_{k_1+\frac{Q}{2}}c^\dag_{-k_1+\frac{Q}{2}}+\frac{1}{2}\Delta_Q(-k)c^\dag_{-k_1+\frac{Q}{2}}c^\dag_{k_1+\frac{Q}{2}}+h.c.
\end{align}
We can write down the BdG Hamiltonian,
\begin{align}
    H_{BdG}=\sum_{k>0}\begin{pmatrix}
        c_{k+\frac{Q}{2}}^\dag&c_{-k+\frac{Q}{2}}
    \end{pmatrix}
    \begin{pmatrix}
        \epsilon_{k+\frac{Q}{2}}&\frac{1}{2}(\Delta_Q(k)-\Delta_Q(-k))\\ \frac{1}{2}(\Delta^*_Q(k)-\Delta^*_Q(-k))&-\epsilon_{-k+\frac{Q}{2}}
    \end{pmatrix}
    \begin{pmatrix}
        c_{k+\frac{Q}{2}}\\c_{-k+\frac{Q}{2}}^\dag
    \end{pmatrix},
\end{align}
We can get the spectrum by diagonalizing the BdG Hamiltonian,
\begin{align}
    E_k=e_0(k)\pm\sqrt{e_1(k)^2+(\frac{1}{2}|\Delta_Q(k)-\Delta_Q(-k)|)^2}=e_0(k)\pm \eta_k,
\end{align}
with $e_0(k)=\frac{1}{2}(\epsilon_{k+\frac{Q}{2}}-\epsilon_{-k+\frac{Q}{2}})$ and $e_1(k)=\frac{1}{2}(\epsilon_{k+\frac{Q}{2}}+\epsilon_{-k+\frac{Q}{2}})$.
The self-consistent equation becomes
\begin{align}
    \Delta_Q(k)=\sum_{k^\prime}V(k^\prime-k)\Lambda_{k_1+\frac{Q}{2},k_2-k_1}\Lambda_{-k_1+\frac{Q}{2},k_1-k_2}\frac{1}{2}\frac{\Delta_Q(k)-\Delta_Q(-k)}{2\eta_k}(\frac{1}{e^{\beta(e_0+\eta_k)}+1}-\frac{1}{e^{\beta(e_0-\eta_k)}+1}),
\end{align}
where the order parameter $\Delta_{\bf Q}({\bf k})$ satisfies $\Delta_{\bf Q}(-{\bf k})=\Delta_{\bf Q}({\bf k})$.
\section{Details on the determination of the critical temperature}
The critical temperature can be calculated from the self-consistent gap equation. We can start from two temperatures $T_a$ and $T_b$, satisfying $T_a\sim 0$, and $\text{max}(|\Delta({\bf k})|)=0$ at $T_b$. For each temperature, we can solve the gap equation iteratively. If $\text{max}(|\Delta({\bf k})|)=0$ at $T_a$, then the critical temperature is $T_c=0$. The critical temperature is defined as $\text{max}(|\Delta({\bf k})|)_{T_c}=\delta\Delta$, where $\delta\Delta$ is the threshold. We can apply the bisection method to solve the critical temperature in the regime $T\in[T_a,T_b]$.

\section{Derivation of the mean-field free energy}
\begin{align}
    H=&\sum_{k}(\epsilon_k-\mu)c^\dagger_kc_k+\frac{1}{2A}\sum_{k_1,k_2}V(k^\prime-k)\Lambda_{k_1+\frac{Q}{2},k_2-k_1}\Lambda_{-k_1+\frac{Q}{2},k_1-k_2}c^\dagger_{k_1+\frac{Q}{2}}c^\dagger_{-k_1+\frac{Q}{2}}c_{-k_2+\frac{Q}{2}}c_{k_2+\frac{Q}{2}}
\end{align}
after the mean-field decomposition, we can get the Hamiltonian
\begin{align}
    H_{MF}=&\sum_{k>0}(e_0+\eta)\gamma_1^\dagger\gamma_1+(e_0-\eta)\gamma_2^\dagger\gamma_2-\frac{1}{2A}\sum_{k,k^\prime}V(k^\prime-k)\Lambda_{k_1+\frac{Q}{2},k_2-k_1}\Lambda_{-k_1+\frac{Q}{2},k_1-k_2}\langle c^\dagger_{k_1+\frac{Q}{2}}c^\dagger_{-k_1+\frac{Q}{2}}\rangle\langle c_{-k_2+\frac{Q}{2}}c_{k_2+\frac{Q}{2}}\rangle\nonumber\\
    +&\sum_{k>0}(\epsilon_{-k+Q/2}-\mu),
\end{align}
the free energy is $F=-T\log{Z}$, where $Z=\int e^{-S}$, finally we can get the free energy
\begin{align}
    F(\Delta({\bf k}))=&-T\sum_{k>0}\log(1+e^{-\beta(e_0\pm \eta_k)})+\sum_{k>0}(\epsilon_{-k+Q/2}-\mu)\nonumber\\
    -&\frac{1}{2A}\sum_{k,k^\prime}V(k^\prime-k)\Lambda_{k_1+\frac{Q}{2},k_2-k_1}\Lambda_{-k_1+\frac{Q}{2},k_1-k_2}\langle c^\dagger_{k_1+\frac{Q}{2}}c^\dagger_{-k_1+\frac{Q}{2}}\rangle\langle c_{-k_2+\frac{Q}{2}}c_{k_2+\frac{Q}{2}}\rangle.
\end{align}
 At $\Delta=0$, the free energy becomes
\begin{align}
    F(0)=-T\sum_{k>0}\log(1+e^{-\beta(\epsilon_{k+{Q}/{2}}-\mu)})-T\sum_{k>0}\log(1+e^{-\beta(-(\epsilon_{-k+{Q}/{2}}-\mu))})-T\sum_{k>0}\log e^{-\beta(\epsilon_{-k+{Q}/{2}}-\mu)},
\end{align}
the last two term together is $-T\log(e^{-\beta({\epsilon_{-k+Q/2}}-\mu)}+1)$
, so $F(0)$ is just the free energy in normal state.
\begin{figure}[h]
    \centering
\includegraphics[width=0.5\textwidth]{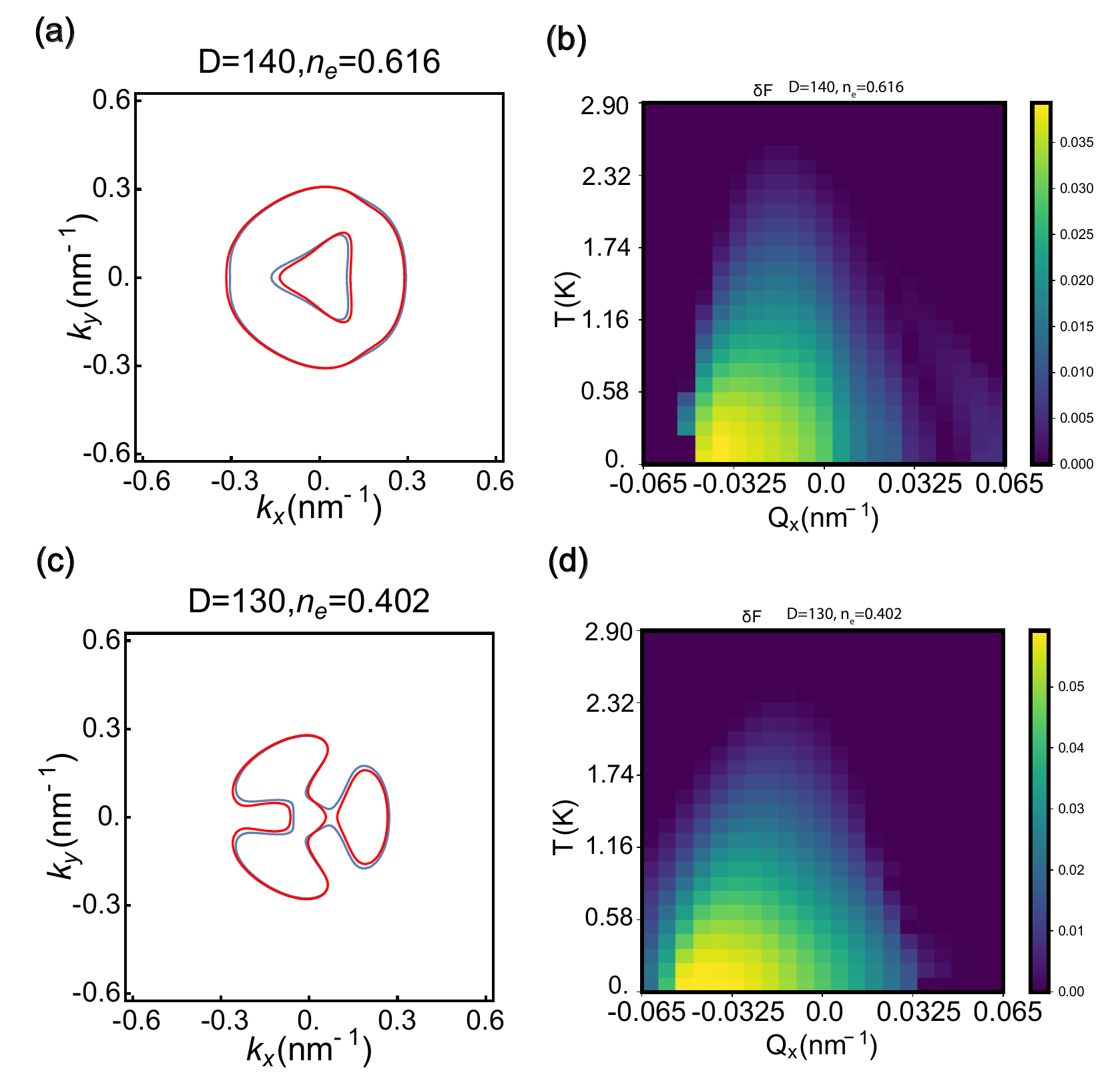}
    \caption{(a) the Fermi surface without (Blue) and with (Red) in plane magnetic field $B_y=10$T at $D=140$meV and $n_e=0.616$. (b) the free energy density for different $Q_x$ and temperature $T$. (c) and (d) the Fermi surface and free energy density for $D=130$meV and $n_e=0.402$. In this calculation, we use $\epsilon=6$.}
    \label{fig:By_10}
\end{figure}

\section{Calculation of the spectral function}
The spectral function is defined as $A(\omega,{\bf k})=-\frac{1}{\pi}\text{Im}(G(\omega,{\bf k}))$, where $G^{-1}(\omega,{\bf k})=\omega+i\eta-H_{BdG}({\bf k})$ is the Green's function. To calculate the Green's at finite temperature, we first calculate the Matsubara Green's function $G^{-1}(i\omega,{\bf k})=i\omega-H_{BdG}({\bf k})$ and in $H_{BdG}$, we use the order parameter calculated at finite temperature. By analytic continuation $i\omega\rightarrow \omega+i\eta$, we can get the Green's function $G^{-1}(\omega,{\bf k})$. The spectral function $A(\omega=0,{\bf k})$ is illustrated in Fig.~\ref{fig:finite_Q}(f).

\begin{figure}[h]
    \centering
\includegraphics[width=0.8\textwidth]{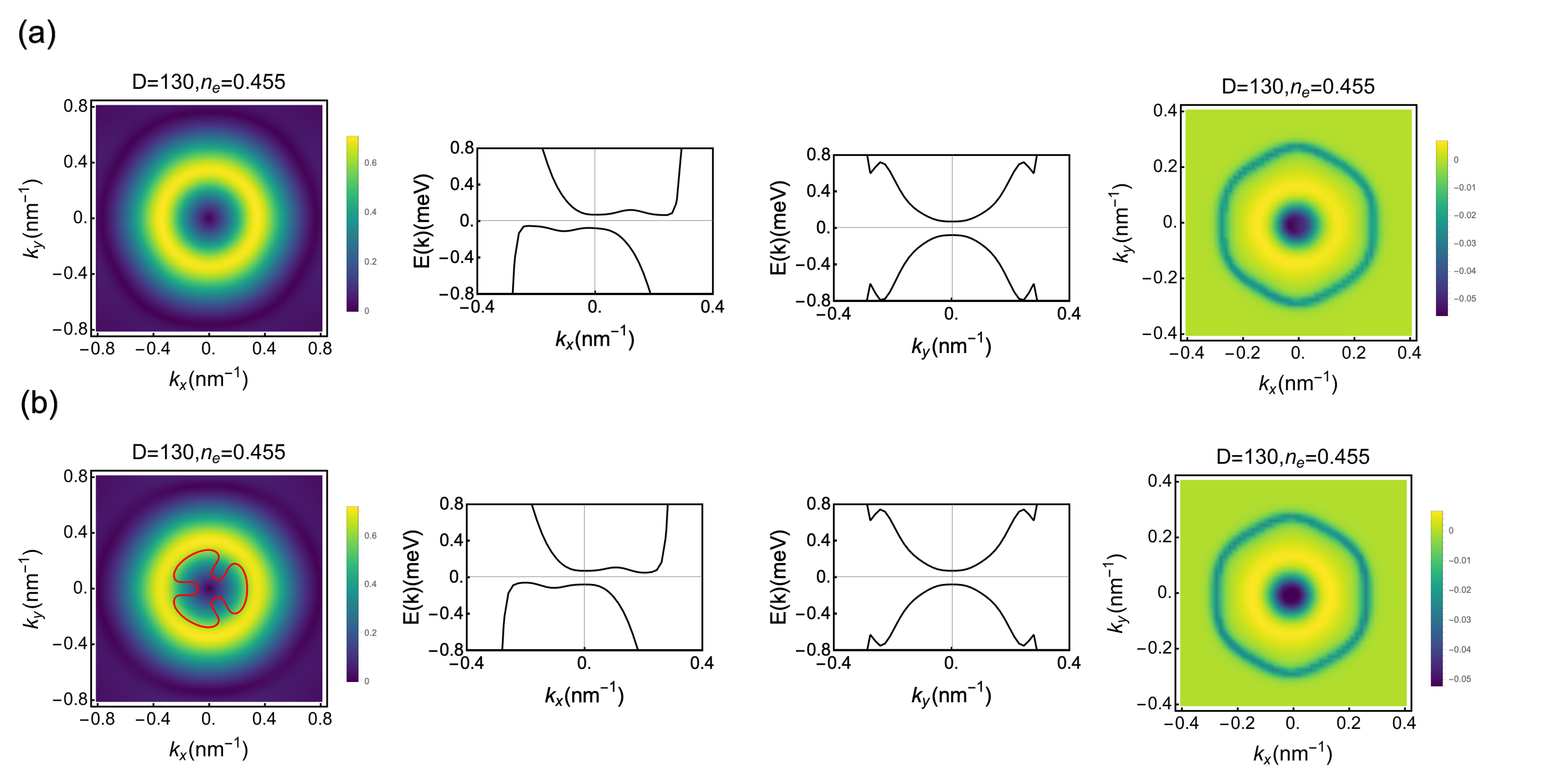}
    \caption{(a) the order parameter $\Delta_{\bf Q}({\bf k})$, the dispersion $E(k_x)$ and $E(k_y)$, the distribution of Berry curvature at optimal ${\bf Q}$ with $Q_x=-0.02nm^{-1}$ for $((D,n_e)=(130,0.455)$, respectively. (b) same plot as (a) for ${\bf Q}=0$.}
    \label{fig:optima_Q_D=130}
\end{figure}

\section{Phase diagram with in plane magnetic field}

In the presence of magnetic field, in Eq.~\ref{eq:free_model} we should replace ${\bf k}$  with ${\bf k}-e{\bf A}$, where ${\bf A}$ is the vector potential corresponding to the applied magnetic field. For an in plane magnetic field we chose the gauge with ${\bf A}=(zB_y,-zB_x,0)$, where $B_x$ and $B_y$ are the components of the magnetic field in the x and y directions, and z denotes the out-of-plane coordinate. This change affects the system's electronic structure and, in particular, alters the shape of the Fermi surface, as shown in Fig.~\ref{fig:By_10}(a) and (c), the Fermi surface becomes more asymmetric in the presence of magnetic field. Specifically, for in-plane magnetic fields, this distortion can cause the Fermi surface to stretch or tilt, allowing for larger pairing momentum $Q_x$ to become energetically favorable. As illustrated in Fig.~\ref{fig:By_10}(b) and (d), the optimal $Q$ increases compared with Fig.~\ref{fig:finite_Q}(a) and (b). The magnetic field breaks the $C_3$ symmetry and thus make the FF superconductor more favorable.

\section{More detailed results}
In Fig.~\ref{fig:optima_Q_D=130}(a), we show the order parameter, the dispersions at optimal ${\bf Q}$ with $Q_x=-0.02nm^{-1}$ for $(D,n_e)=(130,0.455)$, and the distribution of Berry curvature at optimal $Q$. Compared with Fig.~\ref{fig:optima_Q_D=130}(b) at ${\bf Q}=0$, the order parameter and the dispersion are roughly the same. Therefore one can obtain a phase diagram simply using $\mathbf Q=0$.  Although the single particle spectrum is not influenced much by a non-zero $\mathbf Q$, we expect the critical current to lose the $C_3$ symmetry for $\mathbf Q$.

\begin{figure}[h]
    \centering
\includegraphics[width=0.8\textwidth]{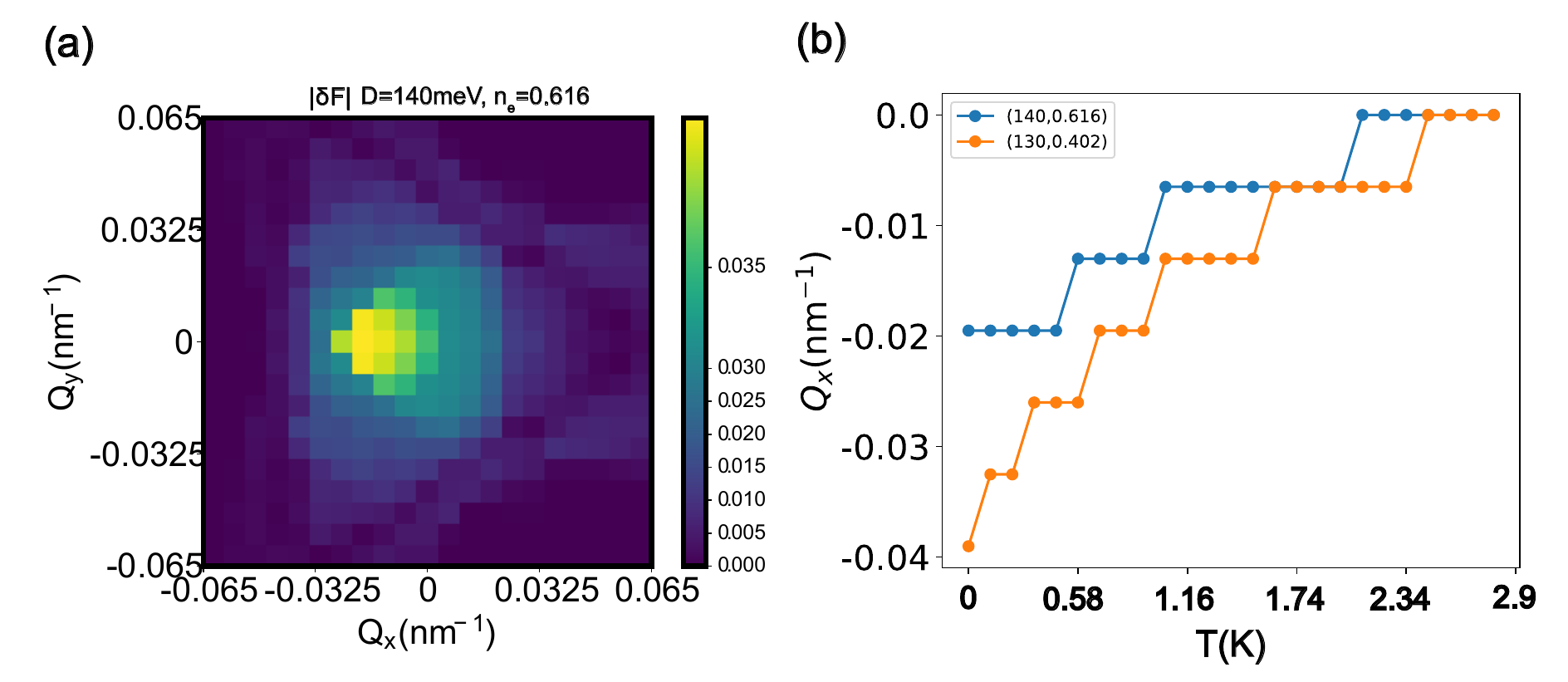}
    \caption{(a) the free energy density dependence of $Q_x$ and $Q_y$ for $(D,n_e)=(140,0.616)$. (b) the temperature dependence of optimal $Q_x$ for $(D,n_e)=(140,0.616)$ and $(D,n_e)=(130,0.402)$. We use $\epsilon=6$ in the calculation.}
    \label{fig:optimal_Q}
\end{figure}

In Fig.~\ref{fig:optimal_Q}, we show the $Q_x$ and $Q_y$ dependence of free energy density and the temperature dependence of optimal $Q_x$. As shown in Fig.~\ref{fig:optimal_Q}(a), the optimal ${\bf Q}$ appears to be along the $Q_y=0$ line. In Fig.~\ref{fig:optimal_Q}(b), we find that at low temperature the optimal $|\mathbf Q|$ gradually decreases with the temperature. Hence it is better to search for the incommensurate FFLO phase at the lowest reachable temperature.

\begin{figure}[h]
    \centering
\includegraphics[width=0.8\textwidth]{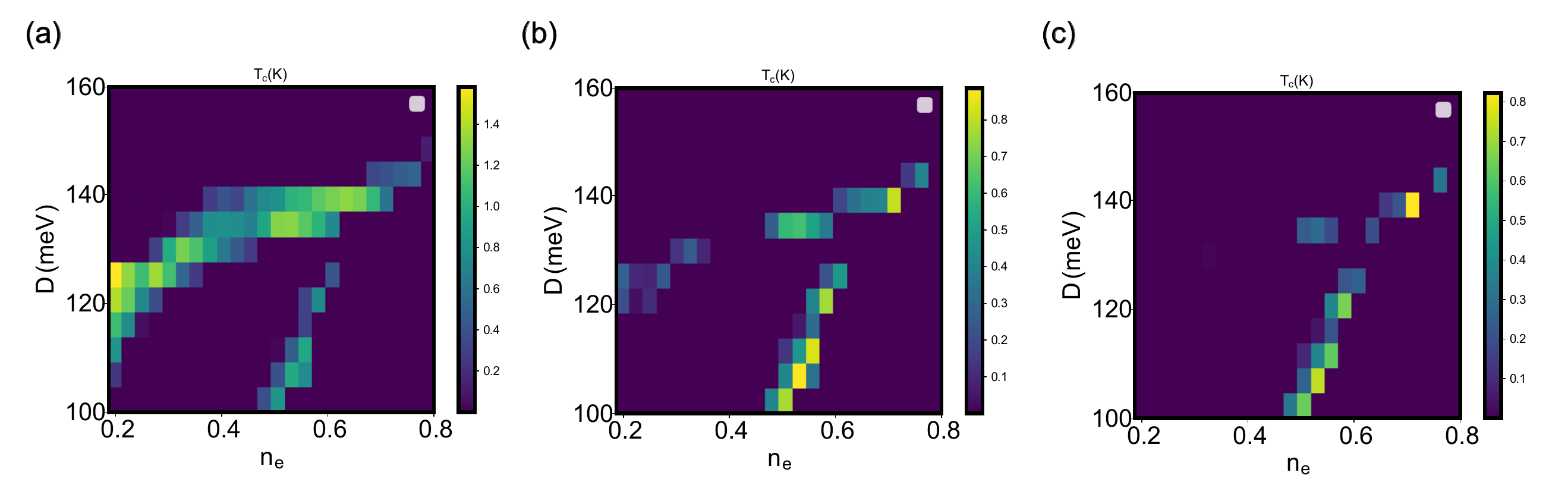}
    \caption{(a), (b), (c) the critical temperature for $\epsilon=10,15,20$, respectively.}
    \label{fig:critical_Tc}
\end{figure}

In the main text we show that the mean field $T_c$ is at order $3$ K for $\epsilon=6$, one order of magnitude larger than the experiment. In Fig.~\ref{fig:critical_Tc}, we show that the mean field $T_c$ gets suppressed at higher $\epsilon$.  A higher $\epsilon$ may be one way to reconcile the disagreement with the quantitative number of experiment. However, we note that our RPA calculation itself is a crude approximation and may overestimate pairing strength.

\begin{figure}[h]
    \centering
\includegraphics[width=0.8\textwidth]{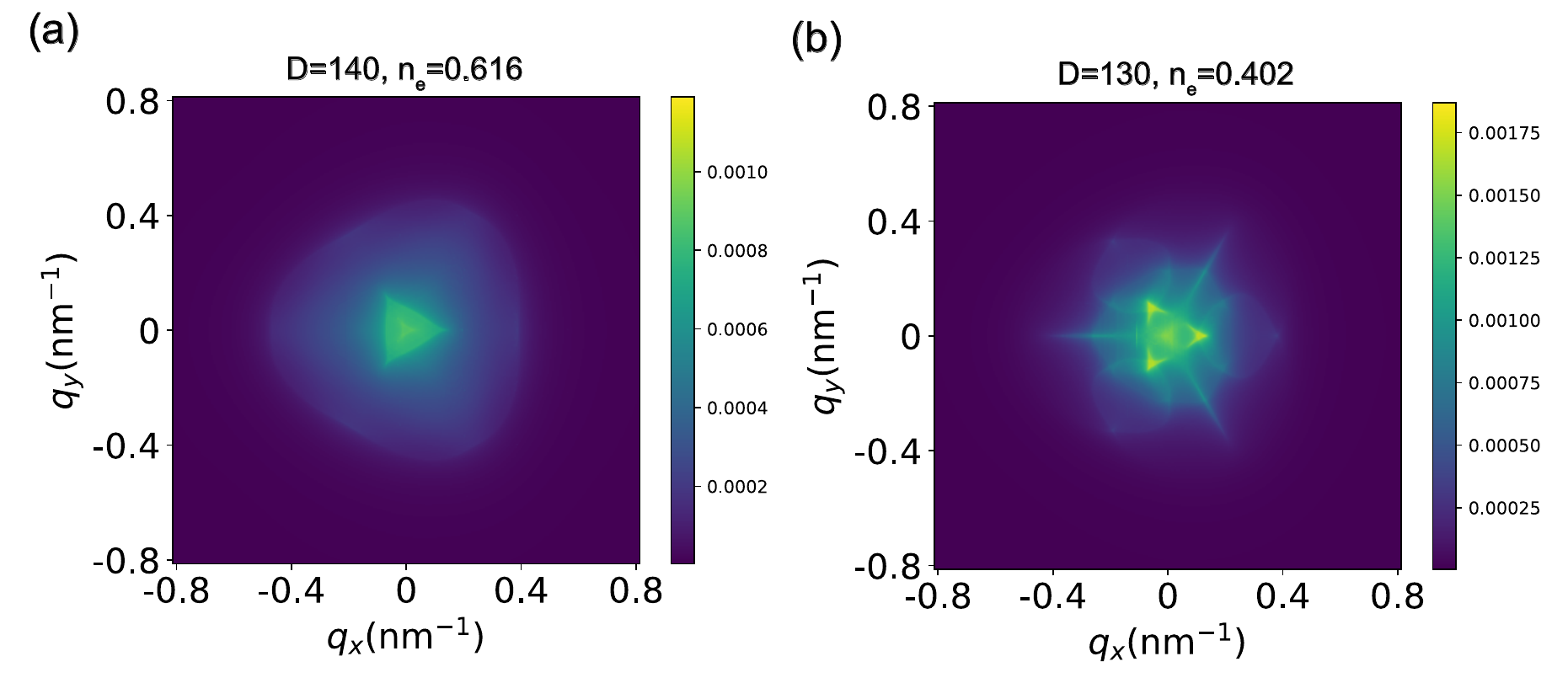}
    \caption{(a), (b) the particle-particle correlation $\Pi^p({\bf q})$ for $(D,n_e)=(140,0.616)$ and $(D,n_e)=(130,0.402)$, respectively.}
    \label{fig:particle_particle}
\end{figure}

The free fermion particle-particle correlation is
\begin{align}
    \Pi^p({\bf q})=\frac{1-n(\epsilon_{\bf k})-n(\epsilon_{-{\bf k}+{\bf q}})}{\epsilon_{\bf k}-\mu+\epsilon_{-{\bf k}+{\bf q}}-\mu},
\end{align}
where $\epsilon_{\bf k}$ is the free fermion dispersion and $n_F$ is the Fermi distribution function. The particle-particle correlation is shown in Fig.~\ref{fig:particle_particle}. We can see the the optimal ${\bf q}$ with maximal $\Pi^p({\bf q})$ does not match the optimal ${\bf Q}$ calculated from the free energy, which indicates the FFLO state is a result of the interplay between the dispersion and interaction.

\end{document}